\documentclass[5p]{elsarticle}

\usepackage[ansinew]{inputenc}
\usepackage{epsfig}
\usepackage{amsfonts}
\usepackage{amssymb}
\usepackage{subfigure}
\usepackage[usenames]{color}
\usepackage{comment}
\usepackage{graphicx}
\usepackage{verbatim}
\usepackage{geometry}
\usepackage{setspace}
\usepackage{epsfig}
\usepackage{subfigure}
\usepackage[usenames]{color}
\usepackage{soul}
\usepackage{comment}
\usepackage{multirow}
\usepackage[cmex10]{amsmath}
\usepackage{bm}
\usepackage{setspace}

\usepackage{algorithm}
\usepackage{algpseudocode}
\usepackage{psfrag}

\begin{document}


\title{Next Generation IEEE 802.11 Wireless Local Area Networks: Current Status, Future Directions and Open Challenges}

\author[bb]{Boris Bellalta\corref{cor1}} 
\ead{boris.bellalta@upf.edu}

\author[lb]{Luciano Bononi}
\ead{luciano.bononi@unibo.it}

\author[rb]{Raffaele Bruno}
\ead{r.bruno@iit.cnr.it}

\author[ak]{Andreas Kassler}
\ead{andreas.kassler@kau.se}

\author[av]{Alexey Vinel}
\ead{alexey.vinel@hh.se}

\address[bb]{Department of Information and Communication Technologies, Universitat Pompeu Fabra, 08018 Barcelona, Spain}
\address[lb]{Department of Computer Science and Engineering, University of Bologna, 40127, Bologna, Italy}
\address[rb]{Institute for Informatics and Telematics (IIT) - Italian National Research Council (CNR), 56124 Pisa, Italy}
\address[ak]{Computer Science Department, Karlstad University, 65188 Karlstad, Sweden}
\address[av]{Centre for Research on Embedded Systems, Halmstad University, }

\begin{abstract}
A new generation of Wireless Local Area Networks (WLANs) will make its appearance in the market in the forthcoming years based on the amendments to the IEEE 802.11 standards that have recently been approved or are under development. Examples of the most expected ones are IEEE 802.11aa (Robust Audio Video Transport Streaming), IEEE 802.11ac (Very-high throughput at $<$ 6GHz), IEEE 802.11af (TV White Spaces) and IEEE 802.11ah (Machine-to-Machine communications) specifications. The aim of this survey is to provide a comprehensive overview of these novel technical features and the related open technical challenges that will drive the future WLAN evolution. In contrast to other IEEE 802.11 surveys, this is a use case oriented study. Specifically, we first describe the three key scenarios in which next-generation WLANs will have to operate. We then review the most relevant amendments for each of these use cases focusing on the additional functionalities and the new technologies they include, such as multi-user MIMO techniques, groupcast communications, dynamic channel bonding, spectrum databases and channel sensing, enhanced power saving mechanisms and efficient small data transmissions. We also discuss the related work to highlight the key issues that must still be addressed. Finally, we review emerging trends that can influence the design of future WLANs, with special focus on software-defined MACs and the internet-working with cellular systems.

\end{abstract}


\maketitle





\section{Introduction} \label{Sec:Intro}
%
\noindent
The IEEE 802.11 standard for Wireless Local Area Networks (WLANs), commonly known as WiFi, is a mature technology with more than 15 years of development and standardisation. The earliest version of the IEEE 802.11 standard was realised in 1997 as a wireless alternative or extension to existing wired LANs using Ethernet technology. However, since its appearance, the IEEE 802.11 specification has continuously evolved to include new technologies and functionalities, and several amendments to the basic IEEE 802.11 standard have been developed. WLANs are currently not only the most common Internet access technology; but they have also expanded across a wide variety of markets, including consumer, mobile and automotive~\cite{2010-commag-survey-wifi}. WLANs are thus widely available everywhere (homes, public hotspots, enterprise environments) and IEEE 802.11-based radio interfaces are found in many types of devices\footnote{According to ABI Research, in 2013 more than two billion IEEE 802.11-enabled devices were shipped.}. 

Several factors have contributed to the success of the IEEE 802.11 family of standards, \emph{interoperability}, \emph{ease of use}, and \emph{flexibility} being among the most important. First, the IEEE 802.11 standards were initially designed to be used within unlicensed spectrum bands, referred to as Industrial Scientific and Medical (ISM) bands. More precisely, most IEEE 802.11 standards work in 2.4 GHz and 5 GHz frequency bands, which are globally available, although local restrictions may apply for some aspects of their use. Thus, anyone can deploy a WLAN in those bands given that a few basic constraints, such as a maximum transmission power, are satisfied. On the downside, this also means that most WLANs are deployed in an uncontrolled fashion with limited or no consideration of interference issues. This has made it especially challenging to guarantee performance bounds and reasonable Quality of Service (QoS) levels. This problem is further exacerbated by \emph{network densification}, i.e., the emerging trend of deploying a large number of base stations in hotspot areas to cope with the increase in traffic demands~\cite{2014-commag-densification}. A second fundamental characteristic of the IEEE 802.11 standards is the adoption of a media access control (MAC) protocol called Carrier Sense Multiple Access with Collision Avoidance (CSMA/CA). The main reason is that IEEE 802.11-based systems are half duplex, i.e., a station cannot carrier-sense/receive while it is sending, and it is hence impossible to detect a collision as in the case of transmissions over twisted copper wires (e.g., using Ethernet). A major advantage of the CSMA/CA method is that channel access procedures are simple and cheap to implement, as they do not impose stringent timing requirements on the radio interface. Furthermore, CSMA/CA protocols are scalable and they provide easy support for mobility and decentralised network architectures, from classical ad hoc networks to emerging people-centric networks~\cite{2014-commag-adhoc,2015-comcom-conti}. On the negative side, CSMA/CA protocols can only provide a best effort transmission service and major efforts have been dedicated to the design of mechanisms for supporting better QoS, such as in the IEEE 802.11e amendment~\cite{2004-wc-survey-qos}. 

The perceived shortcomings of the first WLAN products have driven the evolution of the IEEE 802.11 standards~\cite{2014-comcom-editorial}. In particular, throughput enhancements have been a key priority in the IEEE 802.11 technology development. The key enabler for high-throughput WLANs was the adoption of new physical-layer techniques. The first of these techniques was the orthogonal frequency-division multiplexing (OFDM), which allowed achieving maximum data rates up to 54 Mb/s. However, it is only with the adoption of the IEEE 802.11n amendment in 2009 that the throughput performance of WLANs came close to that of a wired Ethernet network, as a result of the introduction of multiple-input multiple-output (MIMO) technologies~\cite{802.11n-2009}. At the same time, new amendments to the original standard were proposed to foster a more diversified use of WLAN products in various application domains. For instance, the IEEE 802.11p amendment was approved in 2010. This defines enhancements to the IEEE 802.11 standards to support vehicle-to-vehicle (V2V) and vehicle-to-infrastructure (V2I) communication (together referred to as V2X) in the 5.9 GHz band, which is licensed for Intelligent Transportation Systems (ITS)~\cite{802.11p-2010}. Following the same diversification strategy, the IEEE 802.11s amendment was approved in 2011; this described how wireless mesh networks should operate on top of the existing IEEE 802.11 MAC protocol~\cite{802.11s-2011}. This includes the specification of new infrastructure-based elements needed for mesh networking and the routing protocol to establish mesh paths between these elements. In an attempt to consolidate and systematise all the adopted IEEE 802.11 enhancements, the last IEEE 802.11 standard (identified as IEEE 802.11-2012) was finally released to incorporate in an unique specification all the amendments published from 2008 to 2011~\cite{802.11-2012}. 

As pointed out above, the technological development of the WLAN specifications is a continuously evolving process. Thus, while the IEEE 802.11-2012 major revision of the IEEE 802.11 standard was finalised, the IEEE 802.11 working group was also rapidly moving its focus towards next-generation WLANs~\cite{2014-commag-survey-wifi}. Three key drivers were forecasted: $i)$ Machine-to-Machine communications $ii)$ High-Definition Multimedia Communications and $iii)$ ``Spectrum Sharing'' in licensed bands by using cognitive radio technology. Specifically, with the emergence of the \emph{Internet of Things} (IoT) vision, i.e., a world were all sorts of smart objects (ranging from home appliances to small battery powered devices) are connected to the Internet~\cite{2014-comcom-iot}, a low-power WLAN technology is required~\cite{2012-commag-iot-wifi,2014-cl-80211ah}. At the same time, the widespread diffusion of mobile devices with diverse networking and multimedia capabilities, as well as the wide adoption of advanced multimedia applications, is fuelling the growth of \emph{mobile video traffic}, which was already more than half of the global mobile data traffic by the end of 2013~\cite{2013-cisco-report}. Thus, WLANs need specific functions to cope with various multimedia applications, including real-time interactive audio and video, or streaming live/stored audio and video~\cite{Kosek-SzottNSKLST13}. Finally, new regulations for the \emph{unlicensed usage of TV white spaces} are offering new opportunities for additional spectrum utilisation, which can be particularly useful to improve rural coverage of WLANs~\cite{2013-commag-tvspace}. However, cognitive radio mechanisms are required for enabling WLAN communications in TV white spaces. A new generation of amendments is consequently under development or has been completed since 2012 to address these new application requirements. The most relevant are the IEEE 802.11aa (approved in 2012), IEEE 802.11ac (approved in 2013), IEEE 802.11ad (approved in 2012), IEEE 802.11af (approved in 2013), IEEE 802.11ah (in progress, expected for 2016), and IEEE 802.11ax (in progress, expected in 2019), among others\footnote{The association between the IEEE 802.11 amendments and the different use cases is specified in Section~\ref{Sec:Scenarios}.}. 
\begin{table*}[t!!!]
	\centering
	\footnotesize
	\begin{tabular}{llcp{4cm}p{5cm}}
		\hline
		Amendment & Release & Band & Goal &  New features\\
		\hline \hline
		802.11aa-2012 & 2012 & 2.4, 5 GHz & Robust streaming of audio/video streams &  \begin{itemize} \itemsep1pt \parskip0pt \parsep0pt \item Groupcast communication mechanisms \item Intra-access category prioritisation \item stream classification service \item overlapping BSS management
		\end{itemize} \\
		\hline
		802.11ac & 2014 & 5 GHz  & Very high-throughput WLAN in $<$6 GHz band & \begin{itemize} \itemsep1pt \parskip0pt \parsep0pt \item Channel bonding \item Multi-user Downlink MIMO \item Packet aggregation 
		\end{itemize} \\
        \hline		
		802.11af & 2014 & 470-790 MHz (EU) & WLAN in the TV White Space &   \begin{itemize} \itemsep1pt \parskip0pt \parsep0pt \item Geolocation-based spectrum databases \item Channel sensing \item Non-contiguous channel bonding
		\end{itemize} \\
		&& 54-72, 76-88, 174-216, &&\\
		& &470-698, 698-806 MHz (US) &&\\
        \hline		
		802.11ah & 2016 & 902-928 MHz (US) & WLAN in the Sub 1 GHz band & \begin{itemize} \itemsep1pt \parskip0pt \parsep0pt \item Enhanced power saving mechanisms \item Hierarchical station organisation \item Efficient small data transmissions
		\end{itemize} \\
		&& 863-868 MHz (EU)  && \\
		&& 755-787 (China) && \\
		&& 916.5-927.5 MHz (JP) && \\
        \hline		
		802.11ax & 2019  & 2.4, 5 GHz & High efficiency WLANs (HEW) & \begin{itemize} \itemsep1pt \parskip0pt \parsep0pt \item Dynamic channel bonding \item Multi-user Uplink MIMO \item Full-duplex wireless channel
		\end{itemize} \\
		\hline							
	\end{tabular}
	\caption{Summary of the IEEE 802.11 amendments that are reviewed in this survey}\label{Tbl:ListofStandards}
\end{table*}

\begin{table*}[t!!!]
	\centering
	\footnotesize
	\begin{tabular}{llcp{8cm}}
		\hline
		Amendment & Release & Band & Goal \\
		\hline \hline
		802.11ae-2012 & 2012 & 2.4, 5 GHz & Prioritisation of management frames    \\
		802.11ad-2012 & 2012 & 57.05-64 GHz (US) & Very high-throughput WLAN in the 60GHz band  \\
		&& 57-66 GHz (EU)  & \\
		&& 59-62.90 GHz (China) & \\
		&& 57-66 GHz (JP)  &\\
		802.11ai & 2016 & -- & Fast initial link setup  \\
		802.11aj & 2016 & 45, 59-64 GHz & WLAN in the Chinese Milli-Meter Wave frequency bands  \\		
		802.11aq & 2016 & -- & Pre-association discovery (PAD)  \\		
		802.11ak & 2017  & --  & Enhancements for transit links within bridged networks  \\					
		\hline							
	\end{tabular}
	\caption{List of other on-going and upcoming IEEE 802.11 amendments.}\label{Tbl:ListofStandards2}
\end{table*}

In this survey we discuss the most compelling challenges of the new usage models and applications for WLANs that we have identified above. Then, based on those scenarios, we classify and review a selected group of IEEE 802.11 amendments, i.e., IEEE 802.11ac, IEEE 802.11ax, IEEE 802.11aa, IEEE, 802.11ah and IEEE 802.11af, by describing the new technologies and functionalities they introduce to cope with these challenges, such as multi-user MIMO techniques, groupcast communications, dynamic channel bonding, spectrum databases and channel sensing, enhanced power saving mechanisms and efficient small data transmissions. A summary of the main features of these amendments in provided in Table~\ref{Tbl:ListofStandards}. It is important to point out that the IEEE 802.11 specifications do not define all mechanisms, but they typically provide the building blocks and interfaces to allow different manufacturers to implement compatible procedures. Thus, we also provide a detailed review of the main research activities in the various areas and we identify open technical challenges. Finally, we look at emerging new trends for WLANs, with a special interest in Programmable WLANs and LTE-WiFi interworking. Overall, this survey provides a comprehensive overview of the most relevant features in next-generation WLANs, which may be of interest to both researchers and engineers working in the field. For the sake of completeness, in Table~\ref{Tbl:ListofStandards2} we also list the other on-going IEEE 802.11 amendments that have not been analysed in this survey. 

Given the importance of WLANs, other surveys have been published on the IEEE 802.11 standards. Earlier surveys primarily focused on presenting the different classes of proposed MAC protocols~\cite{2000-cst-mac}. A complete overview of the wealth of amendments that have been accepted or were in the process of being standardised before 2010 is provided in~\cite{2010-commag-survey-wifi}. More recently, other surveys have given detailed consideration to specific amendments (e.g., IEEE 802.11s~\cite{2011-cst-survey-80211s}), or classes of similar amendments~\cite{Kosek-SzottNSKLST13,CharfiCK13,2014-commag-survey-wifi}. However, none of the existing surveys follows our use-case oriented approach and covers in such detail all the amendments that we believe will be relevant in coming years. We also include some of the latest advances and related research. 

\begin{figure*}[t]
\centering
\includegraphics[angle=-90,trim=5cm 2cm 4cm 2cm,clip=true,width=10cm]{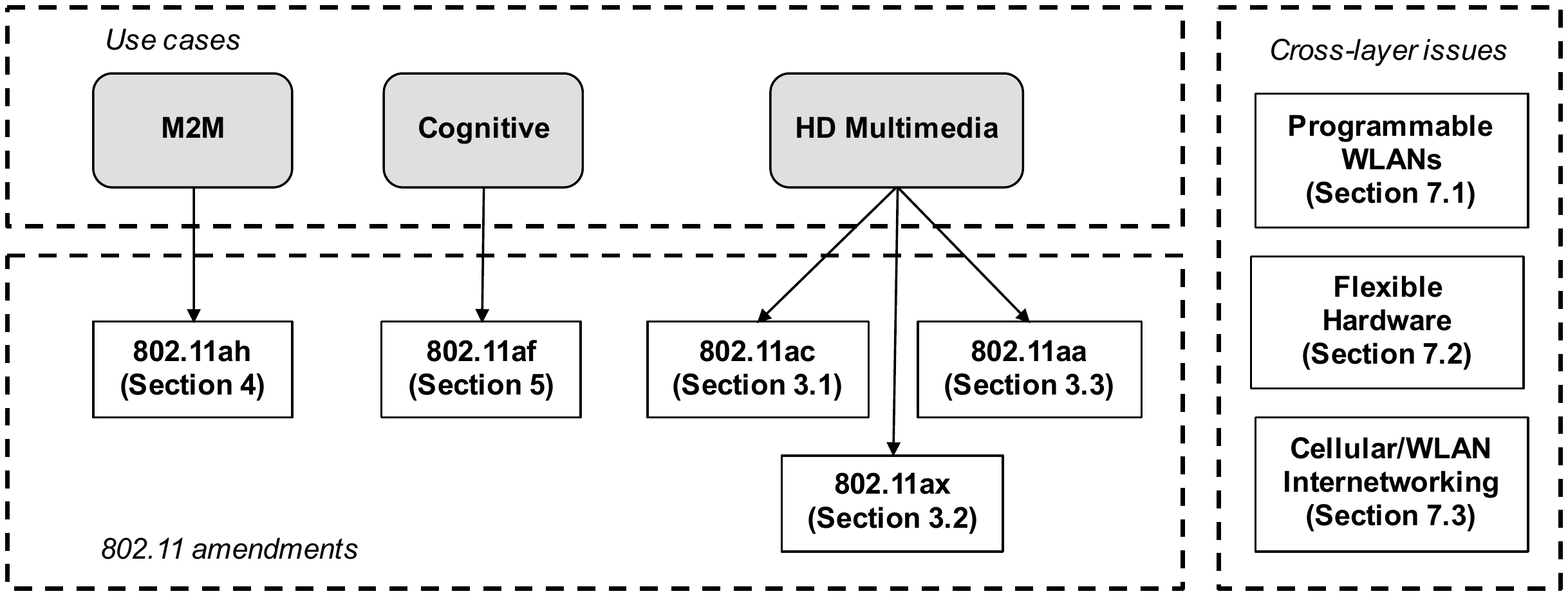}
\caption{Survey organisation}\label{fig:survey_structure}
\end{figure*}
%

The structure of this survey is illustrated in Figure~\ref{fig:survey_structure} and explained in the following. In Section \ref{Sec:Scenarios} we introduce the four key scenarios for WLAN technologies that are considered here. In Section \ref{Sec:HighRate} we focus on high-throughput WLANs, presenting the IEEE 802.11ac, IEEE 802.11ax and IEEE 802.11aa amendments. Section \ref{Sec:M2M} discusses the IEEE 802.11ah amendment to support M2M communications and we review the IEEE 802.11af for WLANs operating in TV white spaces. Finally, Section \ref{Sec:Emerging} presents some emerging trends for future WLANs.

%
%
\section{Future Scenarios \& New Challenges} \label{Sec:Scenarios}
\noindent
WLANs can be found everywhere. They are common in homes, offices, public parks in cities, shops, airports and hotels, among many different places. Today's WLANs are able to provide a fast and reliable wireless access to Internet for browsing the web, exchanging files, chatting, receiving and answering e-mails, and for low-quality real-time audio/video streams, as just a few representative examples of their current usage. This situation is changing rapidly however. The number of persons that use Internet applications and objects that are connected to the Internet is growing every day, proportionally to the number of new applications and services that constantly appear. This clearly results in a steady increase of the Internet traffic. Two representative examples of the change in Internet use are: (i) the high demand for mobile-rich multi-media content, mainly motivated by the use of smart-phones, tablets and other multimedia portable devices; and (ii) the increasing interest in IoT applications driven by the almost ubiquitous existence of devices able to collect data from the environment, ranging from low-power sensor nodes to connected cars. Therefore, WLANs must also evolve to provide effective solutions to these new upcoming scenarios, and the challenges they pose to satisfy their requirements. Four of the key use cases for next-generation WLANs are discussed in the following subsections.

%
\subsection{High-Quality Multimedia Content Delivery}
\noindent
Our new mobile and portable devices are designed to handle rich multimedia contents, including high-definition video and images. Table \ref{Tbl:BHApps} reports the requirements in terms of maximum data rate and latency for some of the most common real-time video applications~\cite{ITU-G114}. Key scenarios in which the support of real-time video transmission is required of course include Internet TV and video streaming.  Similarly, scenarios in which multiple users connect to the same wireless network to request different multi-media content at the same time are increasing every day. However, not all multi-media content is real time. Stored video and image files can also be exchanged between different devices. Those files can have sizes ranging from a few Megabits to several Gigabits, hence requiring a high network transport capacity in order to provide a good Quality-of-Experience to end users. Although video encoding schemes exist that offer substantial video compression efficiency, such as H.264/MPEG-4 AVC~\cite{2007-tcsvt-h264}, WLANs must be able to achieve very high transmission rates and have content-aware mechanisms that are specifically designed for multi-media applications to ensure a satisfactory service for multimedia delivery. The mechanisms that are considered by various IEEE 802.11 standardisation groups to satisfy those requirements are described in Section \ref{Sec:HighRate}, such as group-cast communication protocols, single and multi-user spatial multiplexing and channel bonding among others to make the communication more efficient, and offer higher transmission rates. The reference IEEE 802.11 amendments for high-quality multimedia content delivery are IEEE 802.11aa, IEEE 802.11ac, and IEEE 802.11ax.
\begin{table}[t]
	\centering
	\small
	\begin{tabular}{|l|c| c|}
	\hline
	Type & Max data rate  & Max latency \\
	\hline
	Uncompressed raw video & 1.49~Gbit/s  & 100~ms\\
	uncompressed HDTV  & 150~Mbit/s & 150~ms\\
	Blue-ray Disc & $54$~Mbit/s & 200 ms\\
	MPEG2 HDTV & $19.2$~Mbit/s & 300 ms \\ 	
	MPEG4 HDTV & $8$ to $10$~Mbit/s  & 500 ms\\	
	\hline
	\end{tabular}
	\caption{Performance requirements for different HD streaming applications}\label{Tbl:BHApps}
\end{table}
%
%
\subsection{Machine-to-Machine (M2M) Communications}
\noindent
The almost ubiquitous presence of sensor/actuator devices that are able to interact with the environment has fostered the creation of new services and applications. Concepts such as smart cities and smart grids are being developed on the basis of the existence of those sensor/actuator networks to achieve a more sustainable use of the environmental resources and provide citizens with a higher quality of life~\cite{2013-comcom-smartgrid,etsi-sg}. 

In a classic sense, Wireless Sensor Network (WSN) technologies are used to collect data from spatially distributed sensor nodes and to transmit the data over a multi-hop wireless network to a central sink~\cite{akyildiz2002wireless}. The M2M paradigm is broadening the scope of the WSN concept because it enables networked devices, wireless and/or wired, as well as services, to exchange information or control data seamlessly, without explicit human intervention. Clearly, M2M communications face most of the technical challenges of WSNs. One of the main limitations of WSNs and M2M systems is that the network nodes are usually battery powered or have limited access to power sources. Designing mechanisms and protocols to reduce their power consumption with the goal of extending the network lifetime is therefore crucial for the successful commercial take-up of these kinds of networks. Fortunately, devices in M2M systems typically generate or consume a limited amount of data per unit of time. Thus they can spend a large fraction of their time sleeping. This facilitates energy saving at the cost of additional complexity for the channel access and networking protocols.

Popular wireless protocol standards for M2M communications are Bluetooth, ZigBee and BT-LE~\cite{2014-comcom-iot}. An alternative, promoted by mobile networks, is to connect devices in M2M systems directly to the Internet by using the cellular network infrastructure, for which specific protocols are being developed~\cite{lien2011toward}. WLANs are envisioned as an alternative to both multi-hop WSNs and cellular networks. However, current WLANs are not able to satisfy the minimum requirements for M2M communications~\cite{2012-commag-iot-wifi}. Novel specific power-saving mechanisms are required to support the long periods of inactivity needed by the sensor/actuator devices and to manage the thousands of nodes associated with a single AP. These challenges will be discussed in Section \ref{Sec:M2M}, when presenting the IEEE 802.11ah amendment.
%
%
%
\subsection{Efficient Use of the Spectrum}
\noindent 
The ISM bands are used by several wireless communication technologies, including IEEE 802.11, IEEE 802.15.4 and Long Term Evolution (LTE)-Unlicensed networks. This results in a high spectrum occupancy. Unfortunately, wireless networks operating in the same spectrum region can suffer from mutual interference, which might degrade the performance of all of them. This is exacerbated by the uncontrolled deployment of wireless networks in the ISM band, which is typically very common in urban environments. For example, let us consider a building with several apartments and a WLAN in each one. There would easily be several WLANs operating in overlapping channels and suffering mutual interference \cite{bellalta2015Interactions}. To deal with this issue, it is expected that new APs will increasingly incorporate DCA (Dynamic Channel Allocation) mechanisms to select and update their operating channel at run-time.

An alternative approach to alleviate the spectrum occupancy problem is to move to a different part of the spectrum, even if the new part of the spectrum is occupied by communication systems operating under a license. In that case, WLANs would be the secondary users and therefore must avoid causing interference to the primary users. In recent years, the change from analogue to digital TV broadcast emissions has resulted in a reorganisation of the spectrum at VHF/UHF bands. This reorganisation has shown that there are many empty TV channels, called TV white spaces, that can be used for data communication, especially in rural regions \cite{nekovee2010survey}. Furthermore, WLANs operating in those TV white spaces can take advantage of radio propagation properties in the UHF band to provide large coverage areas. The challenges to be addressed to use CSMA/CA protocols in VHF/UHF bands, as well as how to obtain higher transmission rates when the spectrum is fragmented, will be discussed in Section \ref{Sec:WhiteSpaces}, when presenting the IEEE 802.11af amendment.
%
%
\section{High Performance WLANs for Multimedia Applications} \label{Sec:HighRate}
\begin{figure}[ht!!!!!!!!]
\centering
\includegraphics[angle=-90,trim=2.5cm 0cm 3.5cm 0cm,clip=true,width=9cm]{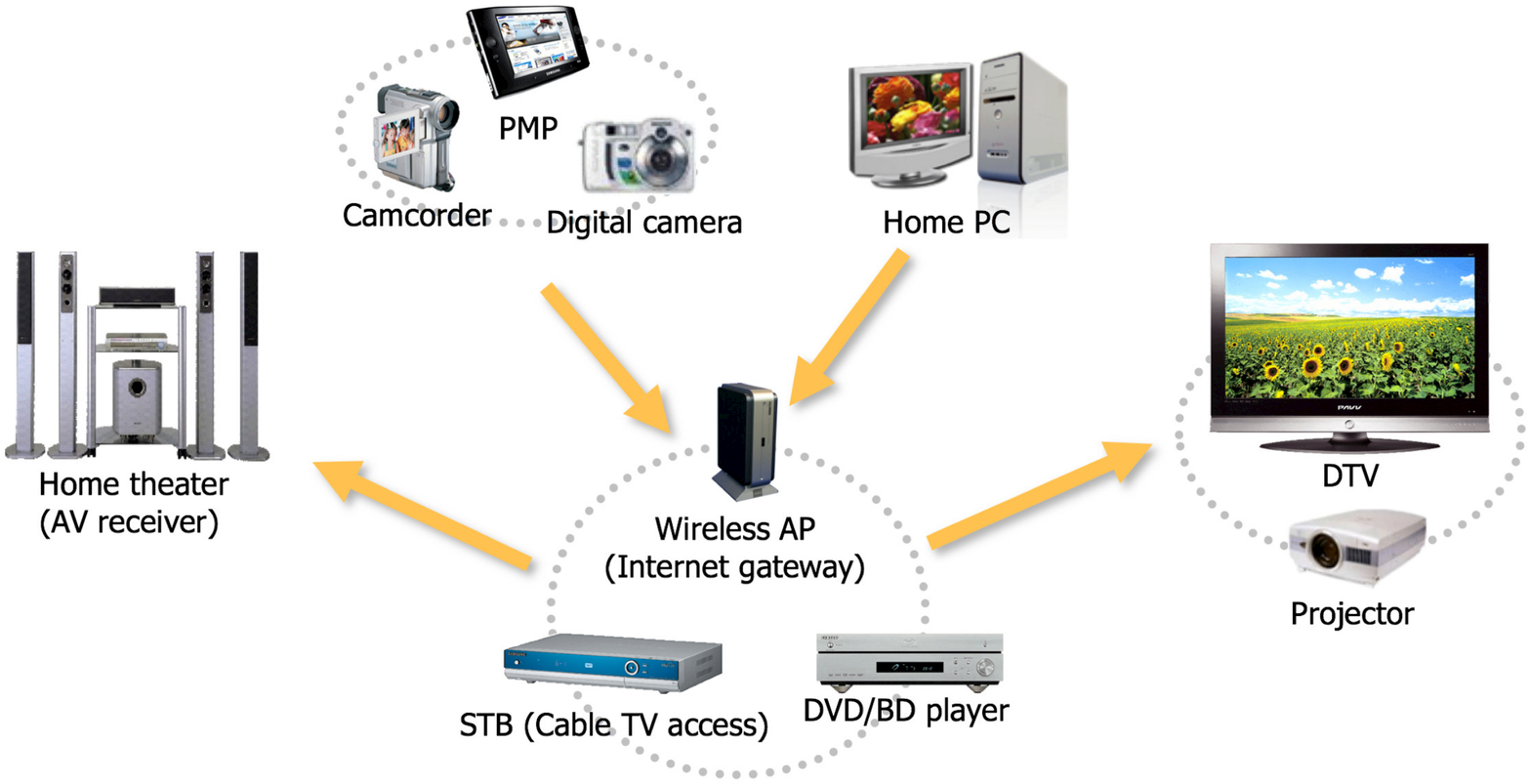}
\caption{High-throughput demanding multimedia devices associated to an IEEE 802.11ac/ax AP.}
\label{Fig:WiFi_Multimedia}
\end{figure}

\noindent
This section reviews the IEEE 802.11ac, IEEE 802.11ax and IEEE 802.11aa amendments. These three amendments target multimedia scenarios by introducing new physical-layer technologies and MAC functionalities to improve the WLAN capacity and QoS provision. Application examples include home scenarios in which an WLAN AP can act as an Internet gateway and wireless media server for home appliances (e.g., IPTV set-top boxes, projectors, game consoles) and content storage devices. A possible use case is illustrated in Figure~\ref{Fig:WiFi_Multimedia}.
%

%
%
\subsection{The IEEE 802.11ac amendment}
\noindent
IEEE 802.11ac \cite{IEEE80211ac} aims to provide users with a throughput close to 1 Gbps, which represents a roughly four-fold increase with respect to IEEE 802.11n~\cite{802.11n-2009}. Compared to IEEE 802.11n, IEEE 802.11ac supports larger channel widths (up to 160 MHz), introduced a new modulation scheme, i.e., a 256-QAM modulation, and downlink multiuser MIMO (DL-MU-MIMO).
%
%
\subsubsection{Novel Features}
\noindent
The most relevant new features included in IEEE 802.11ac are described in the following.
%
%
\paragraph{Channel Bonding}
IEEE 802.11ac enables the use of channel bandwidths of 20, 40, 80 (mandatory) and 160 MHz (optional). Channel bandwidths larger than 20 MHz are created by ``bonding'' (i.e., grouping) a group of consecutive 20 MHz channels, and aim to offer higher transmission rates. 

Two extensions have been proposed in IEEE 802.11ac for the basic DCF (Distributed Coordination Function) access method in order to support channel bonding: (i) the Static Bandwidth Channel Access Protocol (SBCA), which always transmits over the same group of 20 MHz channels, and requires that all sub-channels are idle before starting a packet transmission; and (ii) the Dynamic Bandwidth Channel Access scheme (DBCA), which is able to dynamically adapt the channel width to the instantaneous spectrum availability~\cite{Gong2011ChannelBounding,park2011ieee}. As expected, in dense scenarios the use of DBCA offer a much better performance than SBCA due to adaptability \cite{faridi2015analysis}.

To avoid hidden terminals operating in any of the 20 MHz bonded channels, the IEEE 802.11ac amendment includes extended RTS/CTS frames in order to signal the maximum channel width that can be used at both the transmitter and the receiver. In case the CTS includes a lower channel width than the RTS, the transmitter will adopt it. Similarly to the ACK frames, when the RTS and CTS frames are transmitted, they are duplicated over all the 20 MHz sub-channels used. Note that this enhanced RTS/CTS mechanism is also needed to facilitate the co-existence between IEEE 802.11ac AP and other nearby legacy APs, as the latter could transmit at overlapping times on different sub-channels. The operation and performance of channel bonding in WLANs is thoroughly analysed in~\cite{bellalta2015Interactions}, showing the new interactions between neighbouring WLANs that may appear and their impact in the throughput of each one.

%
%
\paragraph{Downlink Multiuser MIMO}
\noindent
The main novelty introduced by the IEEE 802.11ac amendment compared with the IEEE 802.11n one is the support of MU-MIMO transmissions in the downlink, hence allowing multiple simultaneous transmissions from the AP to different STAs. In the IEEE 802.11ac amendment, the AP can be equipped with a maximum of eight antennas and send up to four spatial streams to two different users, or up to two spatial streams to four different users at the same time. 

When an IEEE 802.11ac AP performs a multi-user transmission it specifies the group of STAs to which that transmission is directed. This information is contained in the new IEEE 802.11ac PHY headers, which are broadcast omni-directionally to all STAs. The way STAs are grouped is decided by the AP after obtaining the channel state information (CSI) feedback from all STAs. To gather the CSI information by the AP, IEEE 802.11ac considers only an explicit channel sounding feedback mechanism called Explicit Compressed FeedBack (ECFB). The channel access is governed by EDCA (Enhanced Distributed Channel Access). At each transmission attempt, the multiple access categories (AC) managed by the AP should contend for the channel medium, as only one AC can be served for each transmission attempt. In the case that the queue associated with the AC that has won the internal contention does not contain packets to enough different destinations to fill all the available spatial streams, it can decide to share the remaining ones with the other ACs.

%
%
\paragraph{Packet Aggregation}
\noindent
To increase the efficiency of each transmission by reducing unnecessary overheads, IEEE 802.11ac allows the transmission of several {MPDU}s aggregated in a single A-MPDU. Then, to acknowledge each MPDU individually a Block ACK packet is used, which contains a bitmap to indicate the correct reception of all included MPDUs. Thus, leveraging on the information contained in the Block ACK, the transmitter is able to selectively retransmit only those MPDUs that have failed instead of the whole A-MPDU. 

%
%
%
\subsubsection{Open Challenges}
\noindent
Since the IEEE 802.11ac amendment has recently been finalised, current research around it should cover two main aspects: a) understanding the performance bounds of IEEE 802.11ac, which entails the development of new models, simulation tools and experimental platforms of IEEE 802.11ac-based WLANs, and b) proposing specific solutions for those aspects that are not defined by the IEEE 802.11ac amendment on purpose, such as the mechanism for creating the groups of STAs for DL-MU-MIMO transmissions, smart packet schedullers able to decide when the use of DL-MU-MIMO outperforms SU-MIMO transmissions, and the implementation of the TXOP sharing feature between several ACs. The results and conclusions obtained in both cases will be very valuable in the development of IEEE 802.11ac technologies, as well as in the conception of the future amendments that will substitute IEEE 802.11ac in four to five years, such as the recently initiated IEEE 802.11ax. 

Following the first mentioned research direction, there are several efforts that have focused on \emph{understanding both theoretical and experimental performance bounds of IEEE 802.11ac}. The maximum downlink throughput that an IEEE 802.11ac AP can achieve when packet aggregation, channel bonding and different spatial stream configurations are considered is presented in \cite{ong2011ieee}. In \cite{zengfirst}, the authors evaluate the IEEE 802.11ac performance experimentally using commodity devices, focusing on the effects that the use of wider channels, the 256-QAM modulation and the number of SU-MIMO spatial streams have in terms of throughput and energy consumption. It is worth mentioning that DL-MU-MIMO was not yet implemented in the equipment they were using, and that feature was therefore not considered. The evaluation of a DL-MU-MIMO implementation for WLANs using the WARP platform is presented in \cite{aryafar2010design}, where a deep evaluation of the potential benefits of DL-MU-MIMO transmissions is done in terms of the location of the receivers, number of users, and user mobility, among other aspects. A solution that combines both packet aggregation and DL-MU-MIMO transmissions is presented in \cite{bellalta2012performance}. Results show the need of properly dimensioning the buffer space to achieve the full potential of such a combination. In \cite{sharon2014mac}, the authors compare the throughput achieved by IEEE 802.11n and IEEE 802.11ac when packet aggregation is used, with and without channel errors. They show that in most cases the packet aggregation mechanism introduced in IEEE 802.11ac outperforms the one in IEEE 802.11n. An analytical model to evaluate the performance of the IEEE 802.11ac TXOP sharing mechanism in DL-MU-MIMO communications is developed in \cite{yazid2014performance}. The main goal of this study is to identify how the TXOP sharing mechanism could improve the system efficiency while achieving channel access fairness among the different ACs. 

How to optimally exploit the new DL-MU-MIMO capabilities provided by IEEE 802.11ac is still an open challenge. First, due to the need of frequent CSI exchanges between STAs and the AP, it is not yet clear in which conditions DL-MU-MIMO outperforms SU-MIMO~\cite{redieteab2012phy,bejaranomute,wangmulti,hiraguri2015survey}, or even whether MU-MIMO does or does not outperform multi-user packet aggregation when the amount of data directed to each destination is not balanced \cite{cha2012performance}. Packet aggregation can be a solution to balance the duration of the multi-user spatial streams as shown in \cite{bellalta2012performance}, although it will always depend on the amount of traffic directed to each destination and the buffer capacity at the AP. In \cite{liao2013performance}, the authors compare different strategies to assign the spatial streams between the available destinations at each transmission in a fully connected mesh network, showing in ideal channel conditions the theoretical benefits of MU-MIMO vs. SU-MIMO. 
     
Closely related to the previous point, a second open challenge is the design of efficient schedulers that consider traffic priorities, the buffer state, the different MIMO strategies, TXOP sharing policies, grouping of STAs and the availability of fresh CSI feedbacks to maximise the throughput and guarantee the required QoS for each active traffic flow. It is important to consider that the availability of updated CSI estimates from all STAs allows the AP to reduce the mutual interference between the transmitted spatial streams, which means lower packet error probabilities and higher transmission rates. However, the overheads for obtaining the CSI from all STAs is large, and increases linearly with the channel sounding rate and the number of STAs. Proposals for reducing the CSI overhead are under development. For example, in \cite{bejaranomute}, the CSI overhead is reduced  by inhibiting the channel sounding whenever possible. Another open problem is how to group the STAs, as the goal is to find groups of STAs with compatible (i.e., orthogonal) channels. In \cite{aboul2013managing}, the authors show the challenges inherent to the group assignment problem, and they propose an heuristic method to solve them. TXOP sharing is considered in \cite{zhu2013enhancing} by presenting two alternative approaches to enhance the considered back-off procedure for the purpose of improving both throughput and fairness. 

A third key challenge for IEEE 802.11ac is to achieve an \emph{efficient use of the spectrum} when several channel widths are used in scenarios \emph{with multiple overlapping WLANs}. Increasing the channel width theoretically allows individual WLANs to achieve a higher throughput. However, the presence of other WLANs in the vicinity also increases the chances of frequency overlapping, which may cause the opposite effect as there appears inter-WLAN contention \cite{bellalta2015Interactions}. Adaptive mechanisms to select the channel centre frequency and the channel width, and MAC protocols to choose the instantaneous channel width used for each transmission are thus required. For instance, in \cite{hanada2013game}, the authors focus on the channel selection problem when WLANs can use multiple channel widths using a game-theoretic framework. In~\cite{2015-comcom-ofdm} a scheme is proposed to enable the communication between nodes with partially overlapping channels, which may provide stronger resilience to channel interferences.


%
\subsection{The IEEE 802.11ax amendment}\label{IEEE80211ax}
\noindent
In 2014 the High Efficiency WLANs (HEW) Task Group \cite{80211ax} initiated the development of a new IEEE 802.11 amendment, called IEEE 802.11ax. The IEEE 802.11ax amendment is expected to be released in 2019, and, to some extent, it will be the IEEE 802.11 response to the expected challenges of future dense WLAN and high-bandwidth demanding scenarios \cite{gong2015advanced,bellalta2015WCM}. 

The open challenges that are considered in the development of the IEEE 802.11ax amendment are to: 

\begin{enumerate}
\renewcommand{\labelenumi}{(\textit{\roman{enumi}})}
    \item Improve the WLANs performance by providing at least a four-fold capacity increase compared to IEEE 802.11ac.
    \item Provide support for dense networks, considering both the existence of multiple overlapping WLANs and many STAs in each of them. Spatial reuse of the transmission resources is a must.
    \item Achieve an efficient use of the transmission resources by minimising the exchange of management and control packets, revisiting the structure of the packets, and improving channel access and retransmission mechanisms, among others aspects. 
    \item Provide backward compatibility with previous amendments. This is achieved by the mandatory transmission of the legacy PHY preamble in all frames, and by keeping EDCA as the basic channel access scheme. 
    \item Introduce effective energy saving mechanisms to minimize the energy consumption.
    \item Support multi-user transmission strategies by further developing MU-MIMO and Orthogonal Frequency Division Multiple Access (OFDMA) capabilities in both downlink and uplink.  
\end{enumerate}

In addition to the aforementioned challenges, next-generation WLANs will have to implement some other functionalities beyond the raw packet transmission and reception. Examples are a fast, efficient and robust handoff between APs in the same administration domain \cite{ong2012performance}, device-to-device communication (D2D) \cite{camps2013device} and coordination of multi-AP networks \cite{suresh2012towards}. In the first case, the IEEE 802.11ai amendment, called Fast Initial Link Setup, is in progress and expected for 2016. Its target is to complete a handoff in less than 100 ms, including new AP discovery, user authentication and configuration. Using D2D communication, we can avoid the use of the AP as a relay, hence improving the overall efficiency as the number of packet transmissions required is reduced. Finally, the virtualisation of network functions adds a new dimension in the management of multiple APs, which in dense scenarios can contribute to notably improving the user experience. We further discuss this last topic in Section \ref{Sec:Emerging}.

Different from the other amendments covered in this survey, the IEEE 802.11ax amendment is just in its initial stages of development, with only very few technical aspects consolidated at this stage. Therefore, in the following subsection, we will overview both the new features and open challenges of the IEEE 802.11ax amendment simultaneously.
%
%
\subsubsection{Novel features \& Open Challenges}
\noindent
The IEEE 801.11ax Task Group is currently working in four areas: \textit{PHY}, \textit{MAC}, \textit{Multi-user}, and \textit{Spatial Reuse}. Next, we will overview some of the topics currently under discussion in the IEEE 802.11 Task Group in each one. 
%
%
\paragraph{PHY layer}
\noindent
The IEEE 802.11ax PHY layer will be an evolution of the IEEE 802.11ac one. The challenges in the design of the IEEE 802.11ax PHY layer are related with the extensions required to support multi-user MU-MIMO and OFDMA transmissions, and Dynamic CCA. Also, improvements in the supported modulation and channel coding techniques will be likely considered to allow for higher transmission rates at lower SNR values. For example, IEEE 802.11ax may consider LDPC (Low-Density Parity Check) coding, which are optional in IEEE 802.11ac, instead of the traditional convolutional codes, as they provide a coding gain of 1-2 dB \cite{gast2013802}. Moreover, the PHY layer may also include some functionalities to support the use of Hybrid ARQ schemes to improve the efficiency of packet retransmissions.
%
%
\paragraph{Medium Access Control}
\noindent
In order to keep backward compatibility with previous IEEE 802.11 amendments, besides a common PHY frame preamble, compatible MAC protocols are required. This means that it is likely that EDCA will be kept as the main channel access technique in the IEEE 802.11ax amendment. Therefore, the most relevant open challenges are related to EDCA extensions to support a large number of STAs, improve traffic differentiation capabilities, improve the energy consumption and provide mechanisms to fairly co-exist with neighboring wireless networks.

To support a large number of contenders with a low collision probability a simple solution is to set a larger backoff contention window compared to the value used in the IEEE 802.11ac amendment for all ACs. To mitigate the extra backoff duration when using larger backoff contention windows, the slot duration can be reduced if the time required to perform the CCA, to switch between reception and transmission modes, and the packet processing delay are reduced. Another approach is to consider decentralised collision-free MAC protocols to enhance the underlying CSMA/CA mechanism in EDCA.  Those MAC protocols are able to build collision-free schedules, thus improving the network efficiency as collisions are reduced, while preserving backward compatibility with the default EDCA implementation. The benefits of decentralised collision-free MAC protocols can be found in \cite{fang2013decentralised}, including a comparative evaluation of several key protocols, including CSMA/ECA \cite{sanabria2013future}. CSMA/ECA is specially relevant since it is fully compatible with EDCA and is able  to adapt to the instantaneous number of contenders. In any case, IEEE 802.11ax WLANs can rely on the IEEE 802.11aa amendment to further improve EDCA traffic differentiation capabilities with intra-AC traffic differentiation and groupcast communication mechanisms, among other features. We overview the IEEE 802.11aa amendment in Section \ref{sec:80211aa}.

IEEE 802.11ax will likely keep the same channel widths that were defined in the IEEE 802.11ac amendment, i.e., $20$, $40$, $80$ and $160$ MHz. However, it is expected that IEEE 802.11ax will extend the use of channel bonding to further improve the spectrum utilisation and the coexistence between neighbouring WLANs. For example, it has been shown in \cite{faridi2015analysis} that the use of dynamic channel bonding provides significant throughput gains in dense scenarios compared with the static approach, while minimizing the inter-WLAN negative interactions \cite{bellalta2015Interactions}. Furthermore, additional mechanisms are required to fully exploit the potentials of using wider channels, such as the use of efficient algorithms to select the position of the primary channel, or even to consider multiple primary channels to increase the number of bonded channel combinations that can be used for transmission.     

The MAC layer in IEEE 802.11ax may work with the PHY layer to implement an efficient Hybrid ARQ mechanism able to retransmit short packets containing only incremental redundancy bits. Opportunistic piggy backing of data packets in ACKs and viceversa may further improve the efficiency of IEEE 802.11ax WLANs by reducing the number of transmissions in a bidirectional data exchange \cite{xiao2005ieee}. Finally, the packet headers can be reduced if shorter STAs identificators are used instead of MAC addresses, and unnecessary fields for the given transmission are not included.
%
%
\paragraph{Multi-User}
\noindent
Multi-user communications will likely be one of the main characteristics of IEEE 802.11ax, as both uplink and downlink MU-MUMO and OFDMA are under consideration. The use of multi-user communication techniques does not necessarily represent a system capacity increase because the available transmission resources may be the same as in the single-user communication case. However, in WLANs, the simultaneous transmission from different users is able to parallelize the large temporal overheads of each transmission (i.e., DIFS, SIFS, ACKs, packet headers, etc.) which can notably improve the WLAN efficiency.  

IEEE 802.11ax will further develop the MU-MIMO capabilities of IEEE 802.11ac by allowing multiple simultaneous transmissions in the uplink, which is known as uplink MU-MIMO. The performance benefits of uplink MU-MIMO have been already extensively studied, showing its benefits but also the requirements and challenges to implement such a solution \cite{liao2014mu}. Similar to DL-MU-MIMO transmissions, an open challenge to enable uplink MU-MIMO is to design a mechanism able to efficiently schedule the users that will transmit simultaneously. In one hand, a pure decentralized approach would be easy to implement with minimal signalling overheads. However, since it requires that all STAs finish their backoff at the same time it may show a very low efficiency, besides that those STAs may not be compatible in terms of their spatial signature. In the other hand, a pure centralized approach requires that the AP has complete CSI and buffer occupancy information from all STAs to select the most suitable group to perform a multi-user transmission. Once a suitable group of STAs is selected by the AP, a "Trigger" frame may be used to notify the group of selected users that can initiate a transmission. This approach guarantees efficient multi-user transmissions but requires some extra overheads to collect all the required information by the AP and signal the selected STAs. In both cases, new multi-user ACKs will be likely introduced by IEEE 802.11ax to acknowledge all transmissions with a single control packet.

Multi-user OFDMA is also in the agenda for IEEE 802.11ax. Using OFDMA, a channel can be split in several sub-channels and assigned to different users. Likely, OFDMA will be implemented in combination with channel bonding, where each of the 20 MHz subchannels can be assigned to a different user, in both downlink and uplink. Besides that, a similar operation as in the multi-user MIMO case is expected, as there are almost the same challenges to solve. A survey of current OFDMA proposals for WLANs is presented in \cite{li2015survey}, showing also how the use of OFDMA is able to significantly improve the WLAN efficiency.

In addition to Multi-user MIMO and OFDMA, the use of Simultaneous Transmit and Receive (STR) techniques, commonly known as full-duplex transmission, have been suggested for IEEE 802.11ax \cite{gong2015advanced,bellalta2015WCM}. Using STR a pair of nodes is able to transmit and receive simultaneously, which theoretically doubles the channel capacity. The challenge is that both the AP and the STA involved in a full-duplex transmission start to transmit at the same time. Therefore, information about full duplex transmissions can be included in control packets or in the PHY headers from the transmission initiator.

%
%
\paragraph{Spatial Reuse}
\noindent
Dense WLAN deployments are necessary to offer a continuous coverage with high transmission rates. To improve both the co-existence with those neighboring networks and the spatial reuse of the spectrum, a WLAN has two options: (i) minimise its area of influence by reducing its transmit power, and (ii) accept higher interference levels by increasing the Clear Channel Assessment (CCA) level. Use of both techniques may increase the number of concurrent transmissions between neighbouring WLANs, and therefore their capacity, although it may also result in the opposite effect since the achievable transmission rates may be negatively affected by the higher interference levels observed, which is the main challenge to be solved. 

Due the high WLAN dynamics, the use of adaptive systems is crucial but challenging as adaptivity requires extra complexity in terms of computing and memory resources, and there are not guarantees that the implemented solution converges due to the decentralized operation of each WLAN. The use of DSC (Dynamic Sensitivity Control) to dynamically adjust the CCA level is one of the aspects currently under discussion in the IEEE 802.11ax Task Group. Initial works evaluating the performance of DSC for IEEE 802.11ax WLANs show a clear improvement on the spatial reuse and the area throughput \cite{afaqui2015evaluation}. Another example of the achievable throughput gains obtained by adapting the CCA can be found in \cite{jamil2014improving}, where the authors show that gains up to 100 \% can be achieved. Moreover, transmit Power Control (TPC) to mitigate interference between WLANs in dense scenarios is studied in \cite{mhatre2007interference}, showing the need of jointly optimising both TPC and CCA to maximise the network performance. 

Finally, sectorization by using beamforming is also under consideration for the development of the IEEE 802.11ax amendment as a potential solution to improve spatial reuse \cite{liu2010pushing}. Using sectorization, only the nodes of a given area are allowed to receive or transmit data, hence reducing the contention between different networks whenever they activate non-overlapping sectors. A challenge here is to coordinate the different neighboring APs when they belong to different administration domains. Decentralized learning approaches may be implemented to find feasible temporal patterns of non-overlapping sectors. 


%
%
%
\subsection{The IEEE 802.11aa amendment} \label{sec:80211aa}
\noindent
As discussed above, legacy IEEE 802.11 standards do not provide robust and efficient delivery of audio/video streaming services. Thus, the IEEE 802.11aa amendment was developed to include new features and additional mechanisms to improve the performance of real-time multi-media content delivery~\cite{80211aa}. Specifically, IEEE 802.11aa addresses the following five shortcomings of previous 802.11 standards~\cite{MaraslisCB12,Kosek-SzottNSKLST13}: 
\begin{enumerate}
\renewcommand{\labelenumi}{(\textit{\roman{enumi}})}
\item the lack of reliable and efficient support for multicast and group communications;
\item the incapacity of applying traffic prioritisation to different multimedia streams or different types of frames from the same stream;
\item the absence of methods for cooperative resource sharing among neighbouring APs; 
\item the lack of mechanisms for graceful degradation of audio/video streaming quality;
\item the non interoperability with existing IEEE 802.1 standards for Audio Video Bridging (AVB).
\end{enumerate}
In the following sections we present in detail the solutions to those problems introduced in the IEEE 802.11aa amendment. We further discuss the research studies that have provided the basis for the IEEE 802.11aa design and we identify the remaining open challenges. 
\subsubsection{Novel features}
\begin{figure}[ht!!!!!!!!]
\centering
\includegraphics[angle=0,trim=0cm 0cm 0cm 0cm,clip=true,width=7cm]{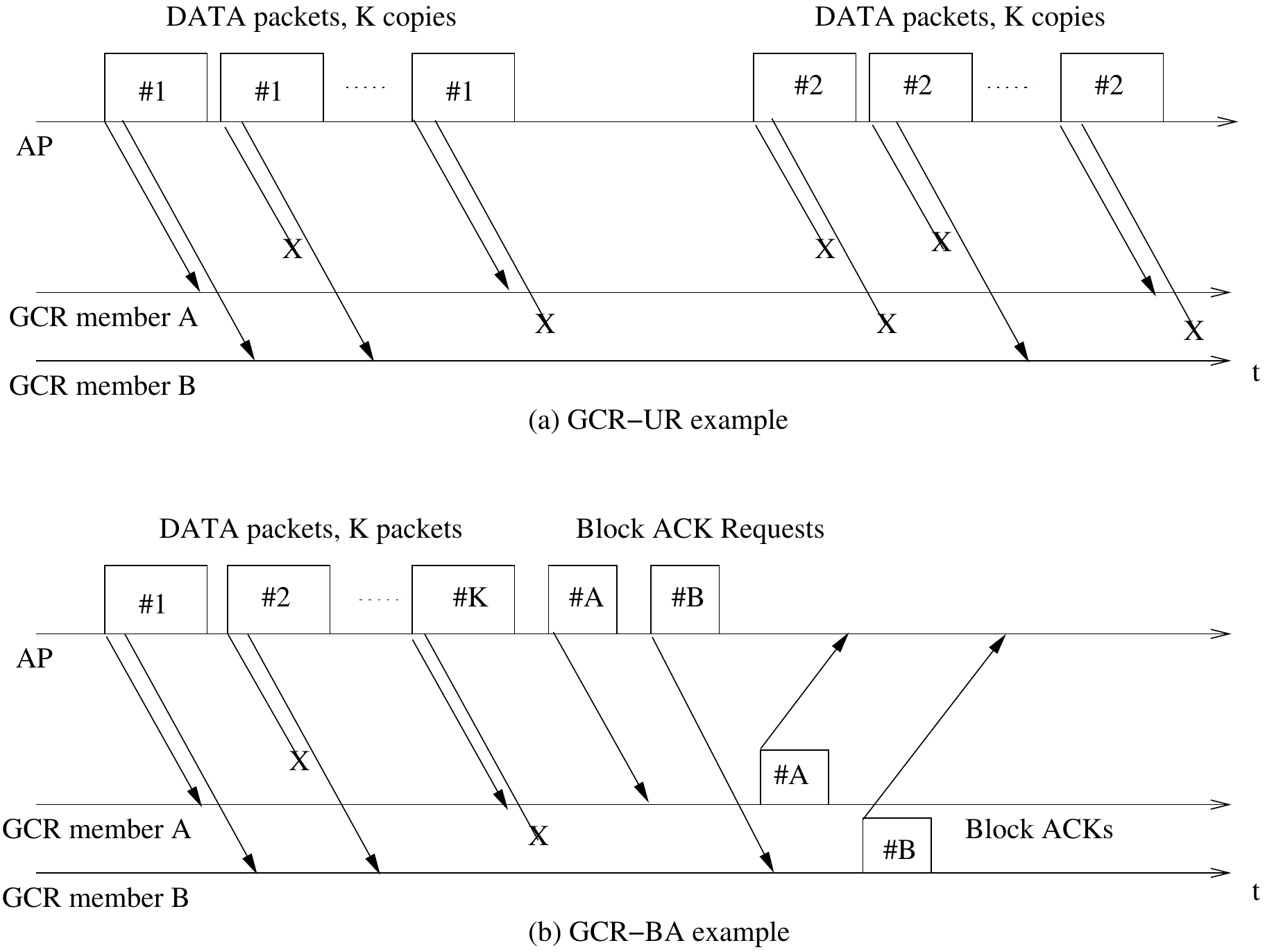}
\caption{GCR service with different retransmission schemes}
\label{Fig:GCR_11aa}
\end{figure}

%
\paragraph{Groupcast communication mechanisms\label{sec:groupcast}}
\noindent
In most audio/video streaming applications a group of clients must receive the same stream simultaneously. A multicast protocol is necessary to avoid that the same content is replicated throughout the network. In wireless networks, multicast transmission can exploit the intrinsic broadcast nature of the wireless channel, i.e., broadcast transmissions from an AP are physically received by all other stations in the same collision domain. However, multicast and broadcast frames in IEEE 802.11 networks are not protected by an acknowledgement mechanism as in the case of unicast frames. Thus, layer-2 multicast transmissions defined by legacy IEEE 802.11 standards are unreliable and not suitable for streaming applications. To partially address this limitation, the Direct Multicast Service (DMS) was first specified in the IEEE 802.11v amendment~\cite{80211v}. Basically, DMS converts multicast streams into unicast streams. In this way, frames destined to a multicast address are individually transmitted as unicast frames to the stations that joined that multicast group. Obviously DMS provides the same reliability as unicast transmission services but the consumed bandwidth increases linearly with the number of group members. To address this scalability issue, IEEE 802.11aa includes the Groupcast with Retries (GCR) service in addition to DMS. Specifically, the GCR service defines new mechanisms and the related management frames for group formation, which allows a set of stations to agree on a shared (non-multicast) address, called the groupcast \emph{concealment} address\footnote{The concealment address protects legacy stations, i.e., GCR-incapable stations, from receiving duplicated group-addressed frames.}. Furthermore, the GCR service specifies two retransmission policies: GCR Unsolicited Retry (GCR-UR) and GCR Block Ack (GCR-BA). When using GCR-UR, the AP can proactively retransmit all groupcast frames a number of times to mitigate the impact of channel errors (see Figure~\ref{Fig:GCR_11aa}.a)), while receivers are not requested to send acknowledgements. Intuitively this approach improves transmission reliability, but it still suffers from scalability issues. In contrast, when GCR-BA is used the AP sends a burst of consecutive groupcast frames and it requests the receivers to reply with a Block ACK frame, which contains a bitmap to positively or negatively acknowledge transmitted frames (see Figure~\ref{Fig:GCR_11aa}.b)). The Block ACK mechanism defined for the GCR-BA service is quite flexible because Block ACK frames can be requested immediately after a transmission burst or after a randomised back-off delay. Furthermore, the AP can request the Block ACK frame to all groupcast recipients or only to a subset of them to reduce overheads and delays. The advantages of the GCR methods over broadcast and DMS have been extensively demonstrated in the literature~\cite{Kosek-SzottNSKLST13,2014-tvt-banchs-multicast}.

\begin{figure*}[t]
\centering
\includegraphics[angle=-90,trim=5cm 1cm 1cm 5cm,clip=true,width=12cm]{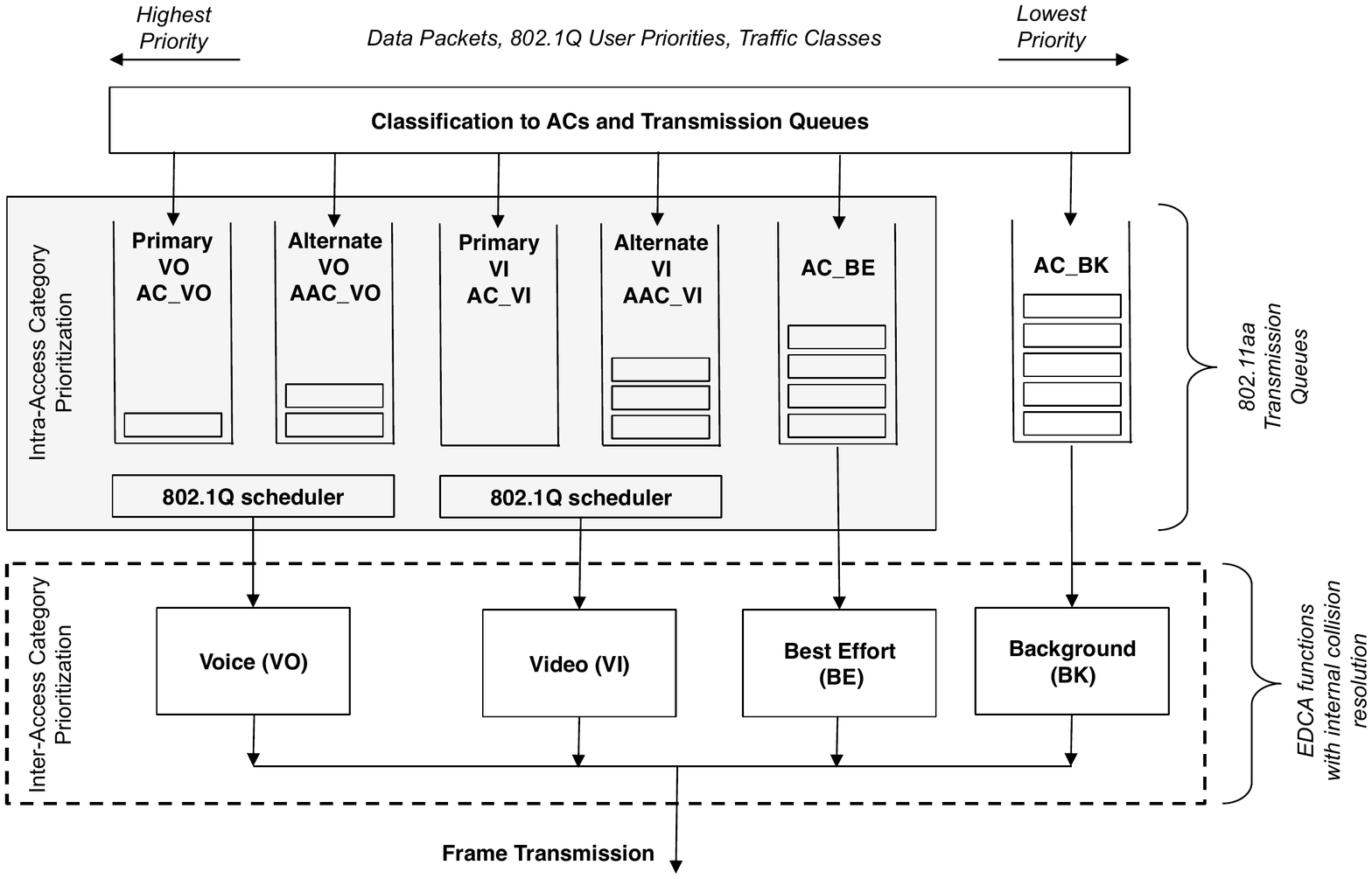}
\caption{Stream classification and inter-AC traffic prioritisation.}\label{fig:inter-access}
\end{figure*}

%
%
%
%
%
%
\paragraph{Intra-access category prioritisation\label{sec:prioritisation}}
\noindent
The IEEE 802.11e amendment only allows traffic differentiation between four different access categories (ACs) that are broadly mapped to four application classes: voice (VO), video (VD), best-effort (BE), and background (BK). However, there is a variety of streaming services, ranging from simple videoconferencing to HD streaming over IPTV systems, which have different QoS requirements (see Table~\ref{Tbl:BHApps}). To provide the ability to differentiate among individual streams, IEEE 802.11aa includes an additional scheduling layer with respect to IEEE 802.11e. IEEE 802.11aa splits each one of the transmission queues associated with voice and video ACs into a primary and an alternate queue. In this way, specialised scheduling rules can be applied to decide which queue to serve when the EDCA function for inter-AC collision resolution grants an access opportunity to voice or video ACs. To facilitate the management of service level agreements, IEEE 802.11aa follows the default mappings between user priority values and traffic types that are defined in the IEEE 802.1D standard~\cite{8021D-2004}. It is then straightforward to further map traffic types onto transmission queues and ACs (see Figure~\ref{fig:inter-access}). Finally, it is important to point out that the intra-AC differentiation functionality can be used to provide more sophisticated traffic differentiation than simple stream prioritisation. For instance, most video applications use Scalable Video Coding (SVC) schemes that enable the partitioning of a video sequence into multiple layers with different qualities and rates~\cite{ElZarkiP94}. Typically, an SVC-based video stream contains a base layer, which provides a basic level of quality, and multiple enhancement layers, which can only be decoded together with the base layer to improve the video quality. Thus, the different layers of the same encoded video steam can be easily mapped to different transmission queues to receive differentiated QoS~\cite{SantosVO12}.
%
%
%
%
%
\paragraph{Stream classification service\label{sec:scs}}
\noindent
The stream classification service (SCS) is an optional service that can be provided by an AP to the associated stations to classify multimedia streams based on arbitrary rules that are established directly by the stations instead of the conventional 802.1D user priorities. To this end the station requesting the use of SCS must initiate an SCS session by sending an SCS request frame to the AP, which contains an identifier for the SCS stream and the descriptor of the classification rule. The AP may accept or reject the requirements specified by the station. Once accepted, the AP must assign all frames that match the classification rule to a specific AC. When intra-access category prioritisation is enabled (see Section~\ref{sec:prioritisation}), matching frames can also be assigned to the primary or alternate transmit queues so that finer grained prioritisation can be applied. Finally, there is also a Drop Eligibility Indicator (DEI) bit in the SCS descriptor that indicates whether frames from this traffic stream can be dropped in the case that there are insufficient resources. Specifically, frames with the DEI bit set to one have a higher probability of being discarded because their maximum number of allowed retries is smaller than the default. Note that how to combine intra-AC queues and frame dropping settings to achieve graceful degradation of the audio/video stream quality in case of bandwidth shortage is beyond the scope of the IEEE 802.11aa specification. 
%
%
%
%
%
%
\paragraph{Overlapping Basic Service Set (OBSS) management\label{sec:obss}}
\noindent
Network densification, i.e., a denser deployment of wireless infrastructure nodes, is one of the key strategies that is used nowadays to easily increase the capacity of wireless systems, even for indoor WLANs~\cite{BhushanLMGBDSPG14}. However, IEEE 802.11 networks have a limited number of orthogonal channels available and, even if optimised frequency planning is applied, it might happen that neighbouring APs are mutually interfering and a station may affect multiple overlapping BSSs. In this case, congestion not only increases but it is also likely to observe an unfair usage of wireless capacity with the channel retained by one AP for long time intervals. This is mainly due to the neighbourhood capture effect, i.e., hidden terminal phenomena among APs~\cite{Benveniste02}. To address this issue, IEEE 802.11aa specifies a new functionality, called Overlapping BSS (OBSS) management, which is based on two new mechanisms. The first defines a set of parameters to quantify the load and interference among neighbouring BSSs, such as medium occupancy fraction, number of admitted audio/video streams, data traffic volumes, and the number of BSSs that are using the same channel as the target one. Note that the traffic load consists of two components: the \emph{allocated} traffic, which is derived on the basis of the TSPEC values of admitted streams\footnote{TSPEC is a traffic specification sent from a QoS capable wireless client that requests a certain amount of network traffic from the AP for the traffic stream it represents.}, and \emph{predicted} traffic, which is evaluated by tracking the maximum value of the allocated EDCA and HCCA traffic over seven-day periods. Once load measurement reports are exchanged among the APs, a second OBSS component is responsible for coordinated admission control procedures on the basis of two suggested sharing schemes: proportional sharing and on-demand sharing. The purpose of both schemes is to keep the total allocated traffic below a maximum value in order to provide some QoS protection to admitted multimedia streams. Finally, IEEE 802.11aa recommends implementing additional OBSS management procedures for channel selection and cooperatively creating HCCA schedules that do not collide.
%
%
%
%
%
%
\paragraph{Interworking with IEEE 802.1AVB\label{sec:avb}}
\noindent
Audio Video Bridging (AVB) is a term commonly used to denote a set of technical standards developed by IEEE to support real-time streaming services with bounded latency through IEEE 802 networks~\cite{8021BA}. This objective is achieved by specifying mechanisms to allow the synchronisation of multiple streams (IEEE 802.1AS~\cite{GarnerH11}) and traffic shaping (IEEE 802.1Qav~\cite{8021Qav}), and to reserve network resources for specific audio/video streams traversing a bridged local area network by using a signalling protocol called the Stream Reservation Protocol (SRP) (IEEE 802.1Qat~\cite{8021Qat}). IEEE 802.11aa integrates the SRP operations with the EDCA admission control procedures. Specifically, the SRP Request/Response messages are encapsulated in the management frames that are used to carry the traffic characteristics and the QoS requirements during admission control. This enables the end-to-end management of resource reservation for QoS guaranteed streams even when one or more IEEE 802.11 links are part of a path from the stream producers (called IEEE 802.1Q talkers) and the stream consumers (called IEEE 802.1Q listeners).
%
%
%
%
%
%
\subsubsection{Open challenges \label{sec:80211aa_research}}
\noindent
In recent years several MAC enhancements have been investigated to improve QoS guarantees for real-time multimedia applications in IEEE 802.11 networks~\cite{CharfiCK13}, and the IEEE 802.11aa standard, which was finalised in 2012, included several of these proposed improvements. Significant research efforts have focused on \emph{improving the transmission reliability of multicasting} by integrating ARQ mechanisms in IEEE 802.11-based multicast transmissions. Modifications to the legacy MAC protocol were proposed in~\cite{KuriK01} to enable the RTS/CTS option in multicast mode and to select one or more multicast receivers (called \emph{leaders}) for acknowledging multicast data packets. However, these enhancements require changes to the standard specifications. The main problems of leader-based ARQ schemes are leader election and the trade-off between scalability and reliability. The authors in~\cite{DujovneT06} propose selecting the multicast recipient operating in the worst channel conditions as the unique leader but this approach may perform poorly in lossy environments. In the Batch mode multicast MAC (BMMM)~\cite{SunHAL02} all multicast recipients are polled by the multicast originator to send individual ACKs, but this scheme is not suitable for large multicast groups. The Enhanced Leader Based Protocol (ELBP) is proposed in~\cite{LyakhovY11} on the basis of multiple ACK-leaders and block acknowledgement techniques. Analytical models are then developed to help select optimal ACK-leaders to meet application QoS requirements. However, the models apply only to saturated traffic while multimedia streams are typically bursty. Another class of reliable multicast protocols relies on busy tones to reduce packet losses due to collisions~\cite{GuptaSL03}, but the additional radio interface needed for the busy tone limits the practicality of such solutions. An alternative approach to avoid collisions of multicast packets is the multicast collision prevention (MCP) scheme~\cite{SantosVBR11}, which is based on the use of a shorter waiting time for transmitting multicast packets. An interesting approach is also proposed in~\cite{2014-comcom-xcor}, to retransmit lost packets using an online linear XOR coding algorithm. However, a modification to the standard MAC protocol is required to enable simultaneous ACK transmissions. In summary, several different methods have been proposed to improve multicast transmission reliability by integrating ARQ schemes into the protocol architecture, but there are not conclusive results on which is the best solution. The choice of the most efficient mechanism depends on a variety of interdependent factors, such as loss ratios, channel congestion, multicast group size, and QoS requirements of multimedia streams. A comprehensive analytical framework is needed to optimise the setting of the parameters for each scheme and to dynamically select the best one.    

As discussed above one main difference between unicast services and multicast services in the legacy IEEE 802.11 standard was the lack of acknowledgements. Another critical difference is that multicast frames must be transmitted using a \emph{fixed} rate in the basic rate set while the transmission rate of unicast frames can be dynamically adapted to the channel and traffic conditions~\cite{AncillottiBC09}. Thus, a group of research papers has investigated the use of \emph{rate adaptation to improve the throughput of multicast services} in IEEE 802.11 networks~\cite{BasalamahSS06,ChoiCSKC07,VillalonCOST07,SantosVO12,LiawTWKHL13}. For instance, the authors in~\cite{BasalamahSS06} propose using RTS frames to allow group members to estimate channel conditions. Each member will then send a dummy CTS frame with a length inversely proportional to channel quality. In this way, the multicast transmitter can use the collision duration to predict the lowest data rate that can be used for group transmissions. The overhead introduced by this mechanism is quite high, however. The solution proposed in~\cite{VillalonCOST07}, called ARSM, also relies on feedback messages sent by the multicast receivers, called multicast response frames, to identify the group member exhibiting the poorest channel conditions. However, in this case a different back off timer is associated with each multicast receiver depending on the SNR of previously received feedback messages in order to prevent collision. An approach similar to the one employed in the Auto Rate Fallback (ARF) protocol, a rate adaptation scheme originally proposed in~\cite{LacageMT04}, is used in~\cite{ChoiCSKC07}. Specifically, the number of successful consecutive transmissions and consecutive transmission failures are used to decide when to increase or decrease the transmission data rate, respectively. A modified ARF scheme is also proposed in~\cite{SantosVO12}, which can be applied to videos that are encoded into two layers, namely the base and enhancement layers. However, \emph{how to integrate rate adaptation with the different retransmission policies} that are defined in IEEE 802.11aa is still an open issue.

One research area that is expected to be crucial in the successful development of IEEE 802.11aa-based products is the \emph{design of efficient scheduling algorithms for supporting voice/video traffic}. Almost all research work in this field has been triggered by the IEEE 802.11e amendment that enhanced the original IEEE 802.11 MAC with two new QoS-aware access mechanisms, i.e., EDCA and HCCA~\cite{80211-2012}. In principle, with a well-designed admission control and scheduling scheme, HCCA is able to provide hard QoS guarantees to traffic flows~\cite{BoggiaCGM07,LeeCR11}. However, HCCA is rarely implemented in IEEE 802.11e-based WLANs owing to its higher complexity and cost concerns. Instead, EDCA is widely adopted. Most papers have thus focused on improving EDCA performance. Many papers have proposed analytical models for various subsets of EDCA functionalities. For instance, a saturation-based performance analysis is conducted in~\cite{Xiao05} by differentiating the minimum back-off window size, the back-off window-increasing factor, and the retransmission limit. The authors of~\cite{ZhuC05,TaoP06} also model AIFS differentiation, while the model in~\cite{XuSVS09} jointly captures all the four EDCA parameters for traffic differentiation. More recent papers have analysed the EDCA performance for non-saturated conditions and for arbitrary buffer sizes~\cite{ZhaoTS13}. The authors in~\cite{2014-comcom-video} have developed an analytical model to predict the QoS levels that can be achieved once a new voice/video flow is introduced in the WLAN. A Kalman filter is proposed in~\cite{2014-comcom-kalman} to obtain estimates on the number of active transmission queues of each Access Category in EDCA. These analytical models can then be exploited to derive the optimal configuration of the EDCA parameters to achieve given performance criteria, or to design admission cotnrol schemes that preserve QoS constraints. For instance, a scheme that assigns contention-window values to achieve pre-defined weighted-fairness goals is proposed in~\cite{ChengCSC06}. A control-theoretic scheme is designed in~\cite{PatrasBS12} with the goal of minimising the video traffic delay. However, most of these solutions rely on non-realistic assumptions about video traffic dynamics. An alternative class of solutions dynamically updates the EDCA parameters based on the observed network conditions. In~\cite{cano2010tuning}, the EDCA parameters are optimised considering a WLAN with rigid and elastic traffic simultaneously, analysing the interactions between both types of traffic. The authors in~\cite{XiaoHL07} specify several bandwidth-sharing mechanisms with guaranteed QoS for voice and video traffic. Measurement-based admission control schemes are proposed in~\cite{XiaoLi04}. A TXOP adaptation method is described in~\cite{LiuZ06} that takes into account video frame sizes and transmit queue lengths. However, the main drawback of these solutions is that they are based on heuristics and hence do not ensure optimal and guaranteed performance. Finally, a third category of research papers tries to improve video performance by designing cross-layer scheduling approaches. Specifically, these works take advantage of multi-layer video encoding to classify the frames according to their importance and assign them to different access categories~\cite{KsentiniNG06}. For instance, the authors in~\cite{HeNL08} define classifiers and waiting time priority schedulers that dynamically change the packet priorities according to end-to-end delay measurements. A disadvantage of this approach however is that an additional adaptation layer may be needed to implement the complex interactions that are typically required between the video coding applications and the MAC layer. We conclude this section by pointing out that existing studies provide the basic design principles and techniques for improving multimedia streaming performance in IEEE 802.11 networks. Still the IEEE 802.11aa standard poses new research challenges that have not been sufficiently explored and that will require innovative solutions. For instance, \emph{scheduling between primary and alternate queues} is still an open research area, as the mapping of individual frames to multiple queues in order to achieve graceful degradation of voice/video quality~\cite{Kosek-SzottNSKLST13}.
%
%
%
%
%
\section{Sensor Networks \& Machine-Type Communications} \label{Sec:M2M}
\noindent
As discussed in Section~\ref{Sec:Scenarios}, M2M communications refer to any communication technology that enables sensor/actuator devices to exchange information and perform actions without the manual assistance of humans. This section reviews the main features currently under consideration in the development of the upcoming IEEE 802.11ah amendment, which targets the main challenges of those networks, such as the energy consumption or the management of many devices.
%
%
\subsection{The IEEE 802.11ah amendment}
\noindent
The IEEE 802.11ah amendment \cite{80211ah} aims to provide WLANs with the ability to both manage a large number of heterogeneous STAs within a single BSS, and minimise the energy consumption of the battery-powered STAs. 

\begin{figure}[t!!!!!!!!]
\centering
\includegraphics[angle=0,trim=0cm 0cm 0cm 0cm,clip=true,width=7cm]{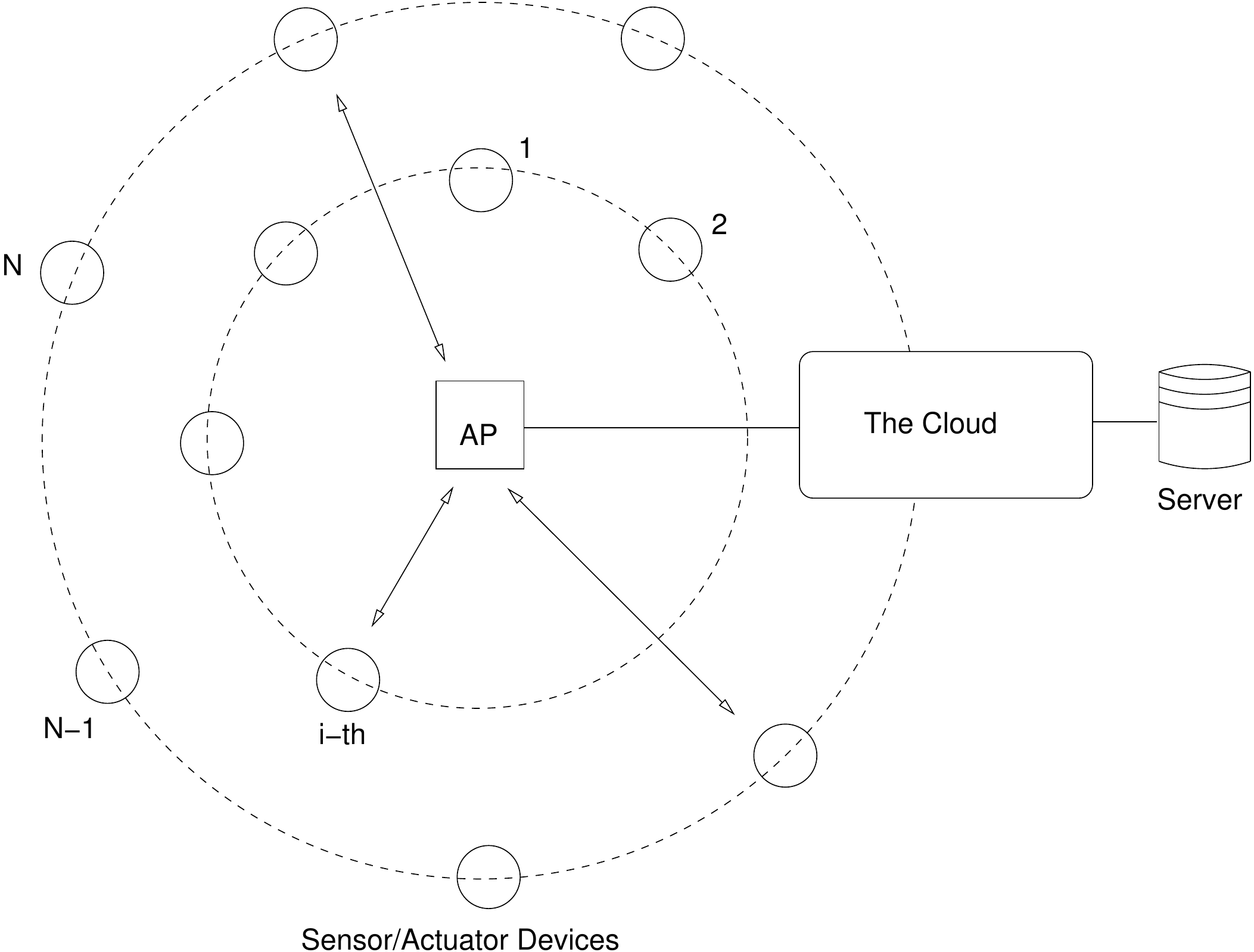}
\caption{WLANs for M2M communications. STAs represent sensor and actuator devices.}
\label{Fig:Wifi_M2M}
\end{figure}

The initial design requirements of the IEEE 802.11ah amendment are detailed in~\cite{adame2014WirMag}; these entail the support of up to 8192 STAs associated with a single AP, the adoption of efficient power saving strategies, a minimum network data rate of 100 Kbps, the operation in the license-exempt Sub 1 GHz band, and a coverage up to 1 km in outdoor areas (see Figure \ref{Fig:Wifi_M2M} for an illustrative example). A preliminary assessment of performance of the IEEE 802.11ah technology, in terms of the number of STAs that can be effectively supported in a WLAN, as well as their energy consumption, is presented in \cite{adame2013capacity}.    

IEEE 802.11ah operates over different sub-1GHz ISM bands depending on country regulations: 863-868 MHz in Europe, 902-928 MHz in the US and 916.5-927.5 MHz in Japan. China, South Korea and Singapore also have specific channelisations. Channel widths of 1 MHz and 2 MHz have been adopted, although 4, 8 and 16 MHz are also supported in some countries. IEEE 802.11ah furthermore proposes new PHY and MAC layers. The IEEE 802.11ah PHY layer can be considered to some extent a sub-1GHz version of the IEEE 802.11ac one. At the physical layer OFDM is the chosen modulation method using 32 or 64 tones/sub-carriers that are spaced by 31.25 kHz. The supported modulations include BPSK, QPSK and from 16 to 256-QAM. A broad range of antenna technologies, ranging from single-user beam-forming to MIMO and DL-MU-MIMO, which was first introduced in the IEEE 802.11ac amendment, are also included in the IEEE 802.11ah specification. Similarly, the IEEE 802.11ah MAC protocol include most of IEEE 802.11 main characteristics, further extending its power saving mechanisms.

%
%
\subsubsection{Novel features}
\noindent
This section gives an overview of how IEEE 802.11ah further extends the IEEE 802.11 PS mechanisms to account for the specific characteristics of resource-constrained sensor and actuator devices, with the aim to offer to the reader a concise vision of the most relevant IEEE 802.11ah features. A more detailed review can be found in \cite{adame2014WirMag}, including a performance assessment of IEEE 802.11ah in several of the key scenarios for M2M communications, such as agriculture and animal monitoring, smart metering, and industrial automation plants. In addition, a detailed survey of the IEEE 802.11ah is reported in \cite{khorov2014survey}, which addresses aspects not considered in \cite{adame2014WirMag} such as the use of sectorisation to avoid overlapping between multiple IEEE 802.11ah WLANs, and the use of relays to directly interconnect IEEE 802.11ah WLANs, among other aspects. Finally, in \cite{park2015ieee}, another overview of the IEEE 802.11ah novel features is presented, introducing both PHY and MAC characteristics with emphasis on the benefits of using the sub-1GHz ISM band in terms of link budget, the location of the OFDM pilots for outdoors operation, the use of bidirectional transmission opportunities, and the new packet formats to minimize overheads, among others. 

%
%
\paragraph{Enhanced Power Saving Mechanism}
\noindent
Power saving (PS) mechanisms for WLANs were already considered in the development of the first IEEE 802.11 standard with the goal of improving the lifetime of battery equipped devices~\cite{2014-comcom-energy}. In PS mode, STAs keep the transceiver in sleeping mode as much time as possible. They periodically wake up to listen to the beacons transmitted by the AP. Those beacons indicate whether an STA has packets waiting for it at the AP. In the positive case, the STA remains awake and requests the delivery of those packets. Otherwise, given that the STA has nothing to receive, it returns to sleep mode until the next beacon is expected.

In the IEEE 802.11ah amendment, time is divided into pages, DTIM (Delivery Traffic Indication Map) periods, TIM (Traffic Indication Map) periods, and slots. DTIM and TIM periods begin with the corresponding DTIM and TIM beacons sent by the AP. The functions of DTIM and TIM beacons are described below:
\begin{enumerate}
\item DTIM Beacons: They inform as to which TIM Groups (i.e., the group of STAs assigned to the same TIM period) have pending packets at the AP. 
\item TIM (Traffic Indication Map) Beacons: Each TIM message informs a TIM Group about which specific STA has pending data in the AP. Between two consecutive DTIMs, there are as many TIM beacons as TIM Groups.
\end{enumerate}
\begin{figure}[t!!!!!!!!]
\centering
\includegraphics[angle=0,trim=0cm 0cm 9cm 0cm,clip=true,width=7cm]{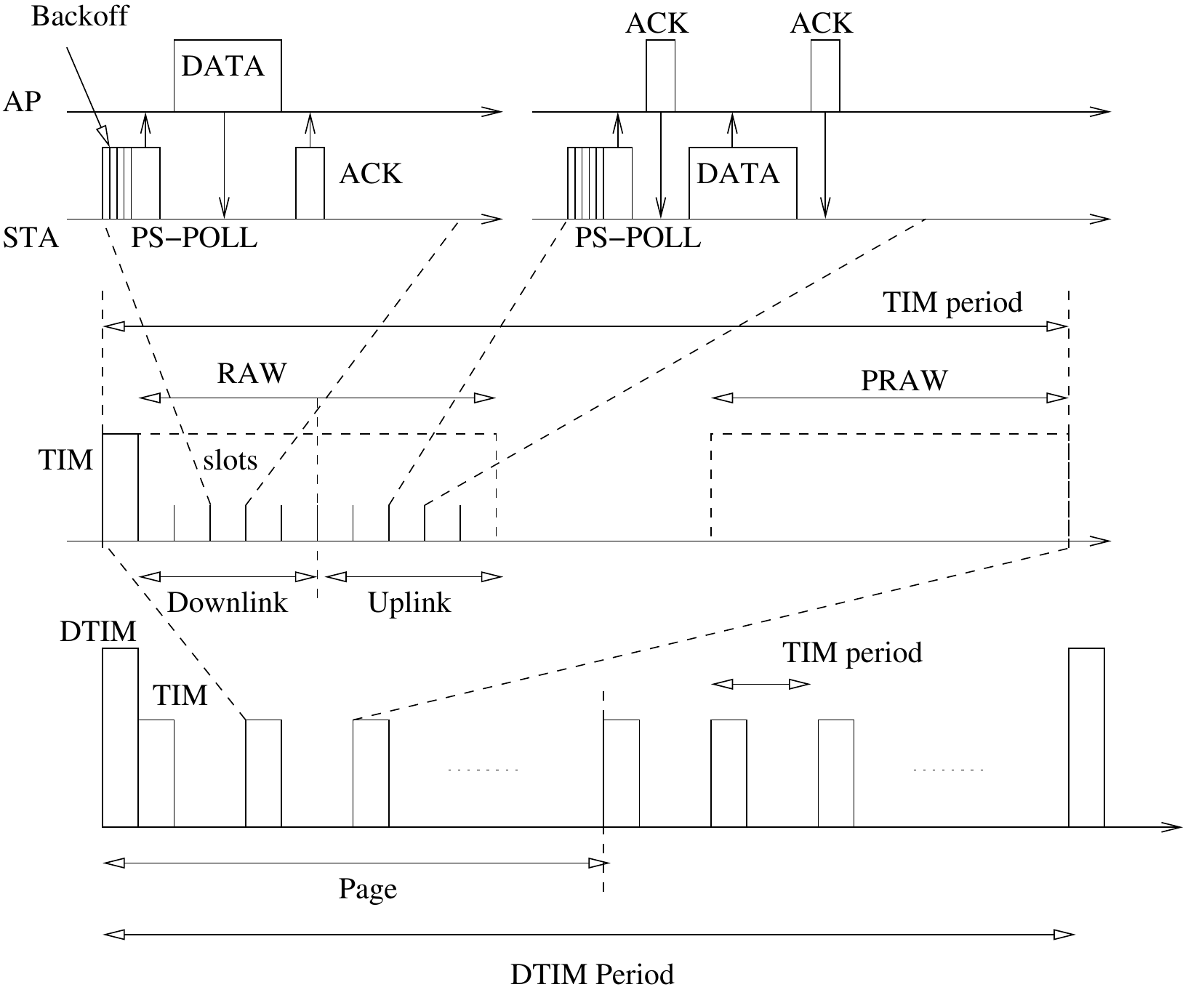}
\caption{IEEE 802.11ah PS mode for TIM-STAs.}
\label{Fig:MAC80211ah}
\end{figure}

Using this DTIM/TIM-based approach, any STA can enter into a power saving state if it does not have packets pending for transmission and one of two conditions is met: 1) it observes in the DTIM beacon that there is no downlink traffic addressed to its TIM Group or 2) it observes in the DTIM beacon that there is some downlink traffic addressed to its TIM Group but where this does not explicitly appear in the TIM beacon. Compared to the preliminary IEEE 802.11 PSM this approach reduces the size of the Traffic Indication Map in each TIM beacon, thus reducing the overhead and the time STAs need to listen and process them. In addition, TIM periods can be organised in pages, which further increases the number of TIM groups between two DTIM beacons.

The temporal organisation of pages, DTIM and TIM periods is reported in Figure \ref{Fig:MAC80211ah}, which shows the distribution of TIM periods in RAW (Restricted Access Window) and PRAW (Periodic RAW). The RAW is a time interval in each TIM period where TIM stations can transmit (see below the different types of STAs defined in IEEE 802.11ah). In addition, it can be divided into several downlink and uplink slots for further granularity. In the downlink, the slots are assigned to a single STA or a group of STAs, while the uplink slots are randomly selected by the STAs with packets ready for transmission. The PRAW is the period of time in each TIM where non-TIM stations can transmit.

%
%
\paragraph{Types of STAs}
\noindent
IEEE 802.11ah supports three types of STAs: TIM, non-TIM and Unscheduled STAs. A TIM station is assigned to a TIM Group (i.e., the group of STAs assigned to the same TIM beacon). Their data transmissions must be performed within a RAW (Restricted Access Window) period. Non-TIM stations do not have to listen to beacons to transmit data. During the association process, non-TIM devices directly negotiate with the AP to obtain a transmission time allocated in a PRAW. The following channel access can either be renegotiated or occurs periodically, depending on the requirements set by the station. Unscheduled stations do not need to listen to any beacons, similar to non-TIM stations. Even inside any restricted access window, unscheduled station can send a poll frame to the AP asking for immediate access to the channel. The response frame indicates an interval (outside both restricted access windows) during which unscheduled stations can access the channel. This procedure is meant for STAs that transmit data very sporadically.
%
%
\paragraph{Hierarchical Station Organisation}
\noindent
To support a large number of STAs and their organisation in pages, DTIM and TIM periods, IEEE 802.11ah assigns to each associated STA a unique identifier of 13 bits, which is called the Association Identifier (AID). Using this new AID, the maximum number of supported STAs is increased from the original 2007 in IEEE 802.11 to 8191 ($= 2^{13}-1$) in IEEE 802.11ah. However, it also allows categorising STAs according to the type of application they are executing, their power level or even their desired QoS by assigning them to different TIM groups. 
%
%
\paragraph{Long Sleeping Periods}
\noindent
IEEE 802.11ah offers TIM, Non-TIM and Unscheduled STAs the possibility to set very long doze times (up to months). The corresponding clock drift produced by such long doze times must be taken into consideration, however, as the higher the time an STA has been asleep, the further in advance it should wake up to avoid possible synchronisation problems with the network.
%
%
\paragraph{Efficient Small Data Transmission}
\noindent
Three new enhancements have been proposed to reduce the overhead when the data packet size is small. First, while IEEE 802.11 contains a 28-byte MAC header, IEEE 802.11ah proposes a short 18-byte version by using AIDs instead of MAC addresses. Second, IEEE 802.11 has defined several null data packet (NDP) frames, which consist only of a PHY header. These frames can be used to create short ACKs, short Block ACKs, short CTSs and short PS-Polls. Finally, a Fast Frame Exchange mechanism has been developed, so that, if an STA has data to transmit, it can notify a successful reception by transmitting its data frame instead of an ACK.
%
%
\paragraph{Sectorisation}
\noindent
Since the PHY layer is based on the IEEE 802.11ac amendment, single and multi-user beam-forming are also supported by IEEE 802.11ah. This allows the transmission of data to multiple STAs simultaneously in the downlink, increasing the system capacity. The use of the beam-forming capability of IEEE 802.11ah APs is also considered as a means to group STAs into different independent antenna sectors with the main goal of reducing interference issues. This would be particularly useful in the case of overlapping with other IEEE 802.11ah WLANs or in the presence of hidden nodes. 
%
%
\subsubsection{Open Challenges}
\noindent
A first open challenge is to \emph{understand the coverage and achievable transmission rates in IEEE 802.11ah WLANs in both indoor and outdoor scenarios}. Propagation models for WLANs working at frequencies lower than 1 GHz are evaluated in \cite{aust2012sub}. The authors compare two path loss propagation models proposed by the IEEE 802.11ah Task Group (one for macro, and one for pico/hotzone deployments)~\cite{TGah-channel-models} with Lee and Hata-Okumura propagation models. Results show that the IEEE 802.11ah channel models underestimate path loss with respect to Lee and Hata models. Moreover, in a comparison with empirical data, it is observed that the IEEE 802.11ah channel models also underestimate the initial loss and the slope of the path-loss function. A new model parameterisation is thus proposed by the authors. In \cite{hazmi2012feasibility}, the feasibility of an IEEE 802.11ah deployment is also evaluated in terms of the achievable range and bitrate, computed on the basis of the link budget using the same path loss propagation models as in \cite{aust2012sub}. Results show that the transmission power limitations in the uplink can limit the overall network performance. Finally, in \cite{sun2013ieee} the authors evaluate the achievable transmission range for the different transmission rates, and the achievable throughput for different combination of transmission power values and transmission rates. Further studies are required due to the heterogeneity of scenarios in which IEEE 802.11ah WLANs can be deployed to characterize obstacles and effective coverage areas, with special attention to outdoors and in presence of mobile nodes.

Single-hop uplink transmissions from distant STAs require high transmission power to reach the AP. Since not all STAs may be able to transmit at the required power, a solution could be the introduction of nodes able to relay transmissions from them. However, the use of relays has to be efficiently harmonised with the operation in power saving mode, allowing temporal periods in which the relays can gather the data from their associated STAs and then periods in which they forward the data to the AP, which is still an open challenge for IEEE 802.11ah WLANs. In case relays are used, data prediction and aggregation techniques could be implemented to make more efficient transmissions.           

Another challenge is the \emph{assessment of the efficiency of IEEE 802.11ah PS mechanisms} and find their optimal configuration. In~\cite{adame2013capacity}, the authors evaluate the capacity of the TIM and page segmentation mechanisms in terms of the maximum number of nodes supported and the energy consumed given a certain network traffic profile. Results confirm that a large number of STAs can be supported with low energy consumption. The impact of the number of nodes in wake mode on both the energy consumption and the delay performance is analysed in~\cite{2014-comcom-energy-delay}. A detailed analysis of the TIM group-based channel access adopted by the IEEE 802.11ah task group is made in \cite{zhengGroupping2013}, where the STAs assigned to each TIM are uniformly distributed between the slots of the RAW. Since only one group of stations is allowed to transmit in each RAW slot, the channel contention is minimised. The number of downlink and uplink slots in a RAW, as well as the duration, is optimised on the basis of the number of active STAs and the network traffic profile in \cite{bel2014cas}. In \cite{park2014enhancement}, the authors propose an algorithm to determine the optimal size of the RAW interval for uplink transmissions based on the estimation of the number of STAs transmitting and the duration of the RAW. Finally, in \cite{liu2013power}, the authors address the contention problems in Smart Grid communication networks with many nodes and periodic traffic when using a IEEE 802.11ah WLAN. It is still a challenge to consider smart systems able to adapt the IEEE 802.11ah parameters to the instantaneous system state in order to save as much energy as possible.

Since the PS mechanisms affect only the operation of TIM STAs, the performance of non-TIM and Unscheduled STAs has not yet been considered in the literature. To identify the scenarios in which the use of non-TIM and Unscheduled STAs are of interest is still an open challenge, as further investigating how non-TIM STAs negotiate over which PRAWs they can use to transmit and receive data. Co-existence issues between the three types of STAs is also an open challenge that requires further studies. 

The \emph{design of strategies to distribute the STAs between all the TIM groups} based on their specific traffic profile, battery level and application priority is another important challenge that is still completely open. EDCA is the default channel access scheme included in the IEEE 802.11ah amendment and provides some basic traffic differentiation capabilities at the packet level. Besides the different access categories (ACs) needing to be renamed and their parameters (Arbitration Inter-Frame Spacing, or AIFS, CW$_{\text{min}}$, TXOP duration) updated to fit the specific traffic profiles of M2M communications, other mechanisms can be implemented. A mechanism to assign the downlink RAW slots when packets from multiple priority levels are waiting for transmission is required, for example. Also, uplink RAW slots are currently selected randomly by the STAs that want to transmit a packet. A mechanism to reserve some slots at each uplink RAW for high priority STAs may therefore be an option, although this may severely degrade the performance of the low priority STAs. Lastly, different repetition patterns for different priority TIM groups may also allow priority STAs assigned to those priority TIM groups to access the channel more often. We expect this challenge will receive much attention in the upcoming years because of the heterogeneity of sensors that will be connected though a single IEEE 802.11ah AP, specially in urban scenarios.


The design of mechanisms to \emph{avoid and resolve congestion situations} when the same event is detected by multiple sensors is another open challenge for IEEE 802.11ah WLANs with many nodes. Similar to what has been discussed for LTE cellular networks \cite{hasan2013random}, the development of mechanisms aware that overload situations may happen is necessary. For instance, a mechanism in which STAs have to wait a random number of DTIM periods before they can start a transmission may be implemented. This approach would distribute the traffic over a longer period of time in overload conditions but, otherwise, would unnecessarily increase the access delay. 


Finally, since IEEE 802.11ah will compete with 4G/5G cellular networks and WSNs to provide M2M connectivity in many different application domains, such as smart cities or e-health, comparative performance studies to determine the strong and weak points of each technology are required, including also aspects such as the cost of the devices and the system reliability.

\section{Cognitive Radio Technology for TV White Spaces}  \label{Sec:WhiteSpaces}
\begin{figure}[h!!!!!!!!]
\centering
\includegraphics[angle=0,trim=0cm 0cm 0cm 0cm,clip=true,width=7cm]{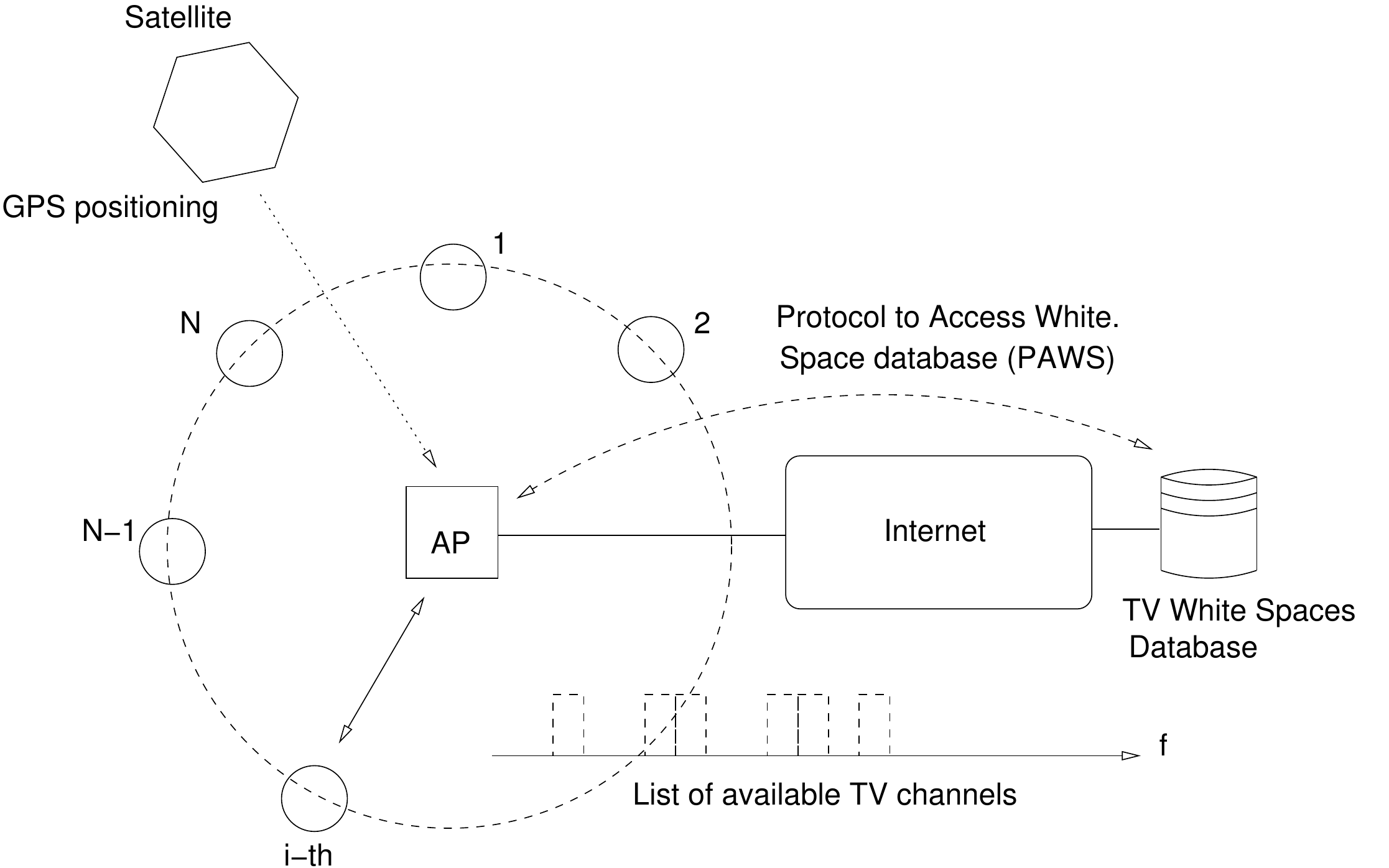}
\caption{WLAN operating in TVWS: basic elements and functionalities.}
\label{Fig:WifiWSpaces}
\end{figure}

\noindent
This section gives an overview of the IEEE 802.11af amendment, introducing its most relevant features and open challenges. Figure~\ref{Fig:WifiWSpaces} shows the basic components and features considered for WLANs operating in the TV White Spaces (TVWS), which include local spectrum sensing by the AP and STAs, and the use of geolocation data bases with information on channel availability.

%
%
\subsection{The IEEE 802.11af amendment}
\noindent
Thanks to the transition from analog TV to digital TV, several VHF and UHF spectrum channels used for decades for analog TV broadcasting are now unused or are under-utilised in many geographical areas~\cite{Davies2011}\cite{Nekovee2009b}\cite{domingo2012white}. This ``digital dividend'' of spectrum has suggested national regulators, such as the FCC in the U.S., and Ofcom in the U.K., to discuss how to reuse these channels for unlicensed devices' communications \cite{FCC2012} \cite{Ofcom2012}. These potentially vacant channels in the VHF and UHF bands are referred to as TVWS  \cite{VandeBeek2012} and include spectrum portions like the 470-790 Mhz in Europe, and non contiguous 54-72, 76-88 MhZ, 174-216 MhZ, 470-698 MhZ and 698-806 MhZ in USA. A snapshot of the TV spectrum occupancy in the city of Barcelona in 2012 is shown in~\cite{sanabria2012spectrum}.

The attractive characteristics of TVWS (not only for WLANs) include the ability to penetrate through walls and other obstacles much more effectively than other widely used spectrum bands, such as the 2.4 and 5.7 GHz ISM bands \cite{Bedogni2014a} \cite{Novlan2010} \cite{Rappaport2001} \cite{Flores2013}. This fact, along with a progressive spectrum scarcity in ISM bands, has suggested the birth of the IEEE 802.11af amendment, published in February 2014 \cite{ieee80211af}, which provides the IEEE 802.11 operational characteristics for TVWS access of unlicensed White Space Devices (WSD). A good summary of IEEE 802.11af can be found in \cite{Flores2013}. The main advantage of operating IEEE 802.11 WLANs in the TVWS comes from an increased coverage range, which can reach up to one kilometre in rural areas and open fields \cite{Simic2011}, and less energy needed to transmit. However, this comes at the cost of an increased interference risk to other WSDs, which creates coexistence problems, and demands new PHY and MAC layers to efficiently support channel access and operations preserving licensed users' devices.

WLANs operated in the TVWS could cover a number of interesting and emerging use cases and scenarios, e.g., Internet access in rural or sparsely populated areas, Smart Grid, sensor aggregation, metering and control, Internet of Things, advanced WLAN operations and TVWS traffic offloading in indoor environments \cite{Nekovee2009a} \cite{Bedogni2013}. Nevertheless, the concept could be realised to create or extend commercial or municipality WiFi services offered to citizens, with coverage at the whole city level realised with reasonably limited network infrastructures \cite{Andersson2011} \cite{Brown2007}.

IEEE 802.11af will use a PHY layer derived from IEEE 802.11ac. It will adopt concepts such as the OFDM, multi-user beam-forming, contiguous and non contiguous channel bonding and packet aggregation. Among the mandatory and innovative behavioural and operational parameters, the most notable one is the channel acquisition support realised through remote geolocation-based spectrum allocation databases, which maintain the channels' availability information in any given area and time of day, providing upon request the list of free channels available for use.

In the following section, we highlight and summarise three main novelties that IEEE 802.11af introduces. In section \ref{sec:geodb} we explain how the access infrastructure to the remote spectrum database is designed and what the requirements are; we discuss the coexistence issues and methodologies adopted in IEEE 802.11af, and we also discuss the novel concept of non-contiguous channel bonding. The following sections also contain an illustration of related works. Open research challenges will be summarised in Section~\ref{sec:openchallenges}.
%
%
%
\subsubsection{Novel features}
\begin{figure}[h!!!!!!!!]
\centering
\includegraphics[angle=0,trim=0cm 0cm 0cm 0cm,clip=true,width=7cm]{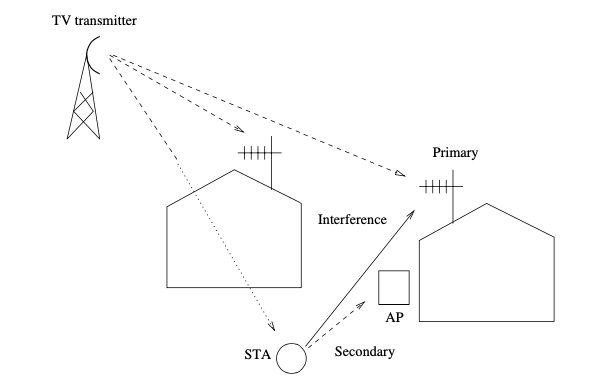}
\caption{Hidden terminal problem in TVWS when the secondary network uses only instantaneous channel sensing to decide whether or not a TV channel is occupied. When the STA is not able to detect the TV signal due to the presence of obstacles, it may initiate a transmission and create interference to the primary users.}
\label{Fig:11afHideenNodeProblem}
\end{figure}

\noindent
This section provides an illustration of novel features introduced and discussed in the path to IEEE 802.11af and the directions provided in related works and standardisation initiatives to resolve classical problems, properly characterised in the new framework of TVWS technologies.
%
%
\paragraph{Channel acquisition: spectrum database and channel sensing}\label{sec:geodb}
Much of the attention in operating in the TVWS is given to the protection of the primary licensed users in the TVWS spectrum band. In general, the primary users were considered as the Digital TV (DTV) broadcasters and receivers. 
In fact receivers would suffer the wasteful effect of collisions during the TV broadcast reception, in the case of secondary users' (SU) transmission interference (Figure \ref{Fig:11afHideenNodeProblem}). 

It is a well-known problem that the definition of a coherent wireless channel status (idle/busy), from the viewpoint of distributed primary users (e.g., DTV receivers) is made difficult by the difference between transmitters and receivers interpretation of signals in a variable time-space collision domain, resulting in hidden/exposed terminals concepts. Solutions proposed to reduce the hidden/exposed terminals included the RTS/CTS mechanisms and beaconing. However, it is difficult to implement solutions that would effectively determine the channel status of TVWS channels based on localised distributed sensing activities performed by multiple secondary users in the same collision domain. On the other hand, the use of RTS/CTS mechanism would not be effective in this case because most of the DTV primary users (and specifically, the receivers) were not originally conceived to transmit signals or implement the RTS/CTS handshake. In conclusion, the actions to identify free (and busy) TVWS channels should not be based on primary users' involvement, but should be almost totally implemented by specifically designed secondary users' methodologies. 

There has been much debate on the methodology that IEEE 802.11af secondary devices should adopt to get knowledge of free channels at a given time on a given target communication area, in an effort to guarantee verified, optimised and reliable TVWS spectrum use. The three main methodologies discussed include: sensing solutions, geolocation databases (DB) and beaconing. Moreover, two approaches are possible, in general: distributed and centralised, with their well known tradeoffs in terms of effectiveness, required infrastructure deployment and coordination overhead. The distributed methodology that is receiving more attention in the literature is based on distributed channel sensing solutions \cite{Akyildiz2011} \cite{Chowdhury2011} \cite{Davies2011}. These solutions include the observation of spectrum use, and possible cooperative aggregation of spectrum sensing information to increase the accuracy of detection and reduce the vulnerability of primary users' transmissions. The spectrum sensing capability is required in all the secondary devices, with a -114~dBm sensitivity. Dynamic transmission power control must be provided and the upper limit on emissions is 100 mW EIRP (20dBm) for portable devices (further limited in the case of adjacent channel use to reduce out-of-band interference). However, database (DB) coordination of TVWS combined with spectrum sensing is considered the most promising and effective technique, compared with spectrum sensing alone. The DB spectrum information has more chances to be effective (i.e., identifying all free channels opportunities) and reliable (i.e., not prone to attacks or erroneous interpretation of channels status). In fact, regulators decided to push for the conservative solution of a remotely accessible, centralised spectrum DB that maintains the information on the availability of all TVWS channels at any given point in time and location (with a target accuracy around 50 m), under their responsibility. For this reason, the spectrum databases have become a mandatory part of many spectrum sharing systems, and the dominant technical solution to support TV white spaces. 

The IEEE 802.11af standard adheres to this vision and also has the role of a common regulatory framework for different spectrum DB implementations, defining a generalised coordination architecture, protocols and interfaces for spectrum queries and local spectrum information management. The protocol explicitly considers out of scope the communication protocol and technology adopted in the background, leaving the freedom to those players responsible for the deployment to select any suitable Internet-based and local access network technology. Secondary devices that need to access the TVWS spectrum should get information on which channels they can effectively use, and which parameters to adopt, without impacting primary users' transmissions (and receptions), by querying the available spectrum DB. The query must contain the transmission characteristics and accurate geographical position of the secondary device, which can be determined by means of a geo-positioning system (GPS), regularly updated in the case of movement (or manually set for static devices). The IEEE 802.11af reference system architecture for DB spectrum access is composed of multiple Geolocation-based Spectrum information Databases (GDBs) entities connected via Internet to Registered Location Secure Servers (RLSS). RLSSs work as local proxies of the GDB for a localised group of Basic Service Sets (BSS). RLSS are connected via a secure protocol architecture with the equivalent of multiple Access Points (AP) in different BSS, realising a trusted DB infrastructure. APs locally coordinate the exchange of information and channel access management between the GDB (via RLSS) and the secondary users' end stations (STA).

Mode I devices are those under the control of a device that employs geo-location database access, while Mode II devices are those employing geo-location database access by themselves. The information exchange provided between the GDB and secondary user STA can be provided in both an open-loop (e.g., adopted by FCC) and a closed-loop (e.g., adopted by ETSI) implementation. In an open loop implementation daily spectrum availability information is provided by the GDB and no feedback on the spectrum information received is provided by STAs; thus the system approach for spectrum access is much more conservative, leading to low channels' utilisation potential. In a closed loop implementation the STA can provide feedback to the GDB, and there are more communication overheads due to the high granularity of updates, however, the system is more effective and reliable in the exploitation of TVWS on behalf of STAs. The typical information provided by the GDB includes 1) the updated White Space Maps (WSM) of frequencies allowed for secondary use at the time/space of the querying STA, and 2) the device-dependent power limitations for transmission (in general, conservative and accurate enough to avoid relevant interference effects on primary users identified in the area). 

The basic IEEE 802.11af mechanisms regulating the communication between STA and GDB can be found in \cite{Flores2013} and \cite{ieee80211af}.

In parallel with IEEE 802.11af, standards such as the IEEE 1900.6 have been created that consider interfaces and data structures supporting spectrum sensing information exchange, applicable to spectrum sensing (and particularly distributed spectrum sensing) scenarios. In particular, in October 2014 the IEEE 1900.6 WG  initiated a project for a new standard, called IEEE 1900.6b \cite{ieee1900.6}, concerning the use of spectrum sensing information to support and optimise the effectiveness, reliability and robustness of spectrum database solutions. The aim is to enhance the performance and capabilities of spectrum databases through the use of spectrum sensing information. 
%
%
\paragraph{Co-existence}
\label{sec:coex}
Support for co-existence mechanisms so that multiple technologies can effectively utilise the TVWS spectrum is important. 
Self co-existence between network devices of a common technology (e.g. deployed by different operators in the same area), and co-existence among different technologies, are relevant topic of research for Cognitive Radio (CR) systems, and specifically for IEEE 802.11af. Many solutions appeared on the research scene, but no one was so far finalized as the target solution for IEEE 802.11af. Specific standardisation has been started to regulate coexistence between wireless standards of unlicensed devices, including the IEEE 802.19.1 \cite{ieee802.19.1}.  The purpose of the IEEE 802.19.1 standard is to enable the family of IEEE 802 Wireless Standards to most effectively use TV White Space by providing standard coexistence methods among dissimilar or independently operated TVWS devices. Early examples of generalized coexistence mechanisms included Dynamic Frequency Selection (DFS), Transmission Power Control (TPC), listen before talk (e.g., for contention based IEEE 802.11, 802.15), time division multiplexing (also among different techniques such as the IEEE 802.16, 802.20, 802.22), and Message-based Spectrum Contention (that is, beaconing messages that carry coexistence information). Opportune metrics must be defined to assess the measurable coexistence achieved among different technologies: as an example, the hidden node probability for a target scenario, or the estimate of percentage variation in normalised network throughput and latency (before and during the SU transmissions). On the other hand, a centralised coexistence control mechanism could be effectively realised by a central manager (or coexistence-DB, like the GDB) in critical scenarios. To this end, IEEE 1900.4  \cite{ieee1900.4} (a standard for heterogeneous networks in dynamic spectrum context, part of IEEE Standards Coordinating Committee 41) aims to standardise the overall system architecture and information exchange between the network and mobile devices, which will allow these elements to optimally choose from available radio resources.

There are three possible classification of co-existence architectures and inter-network coordination channels for CR systems: centralized, coordinated and autonomous co-existence mechanisms \cite{bian2014}.
In \emph{centralized co-existence schemes}, the co-existence is administered by a central entity (e.g like in IEEE 802.19.1). This solution could be applied in both homogeneous IEEE 802.11af systems (e.g. incarnated by the centralized spectrum DB entity) and in heterogeneous inter-networks, without requiring modifications to existing standards, under the assumption that all the involved co-existing network entities would adhere to the same centralized coordination scheme and inter-network coordination protocol. On the other hand, this assumption could be hard to satisfy in some practical scenarios. The centralized schemes are effective in providing minimization of inter-network interference, based on the availability and quality of centralized co-existence information. In \emph{coordinated co-existence schemes}, the centralized entity could be present, but not taking decisions, in general. The central entity could provide a co-existence DB with required information which can be used via in-band and out-band signalling for co-existence decisions taken by the cluster heads of many co-existing inter-networks, for implementing a proper common coordination protocol (e.g. out-band busy tone signalling). These are often hybrid solutions realizing a compromise between effectiveness and suitability. In \emph{autonomous co-existence schemes}, all the decisions and coordination are implemented in a distributed way by the involved network entities. In these schemes, many policies can be adopted to realize a sufficient level of co-existence, under a best effort approach. Solutions that could be implemented include the use of busy tones, beaconing and other signalling protocols in dedicated control channels, dynamic distributed frequency selection schemes, listen before transmit, token-based and dynamic reservation schemes. All the above mentioned solutions have goods and bad, and a general illustration can be found in \cite{bian2014}.
Other proactive co-existence techniques try to early detect and recover/mitigate co-existence issues with relaxed inter-network coordination, e.g. realized via spectrum sensing and interference avoidance/suppression techniques.

It must be clear that different aims exist, regarding the co-existence, for primary TVWS users protection against secondary users, and for secondary devices mutual interference avoidance. Many co-existence problems have been analysed in the literature for TVWS technologies, in particular between the primary (licensed) and secondary (unlicensed) devices. The coexistence management in TVWS is complex because primary users of DVB transmissions are intended to be the pure receivers, rather than the broadcasters. In other words, potential protected users could be everywhere, and the hidden terminal problem could arise for these systems. The co-existence problem is also complicated by many factors that cause asymmetry and dynamicity in the TVWS, such as the mobility, variable density, power asymmetry, and heterogeneous MAC/PHY layers. TVWS co-existence problems originate between the IEEE 802.11af technologies and the IEEE 802.22 technologies for wireless regional area networks (WRAN). Results of analysis of TVWS usage in Europe show that white spaces are typically present and fragmented. They are typically more abundant in rural areas, where larger contiguous blocks of unused channels are available, due to broadcast network planning giving priorities linked to population density. However, the exploitation of TVWS in urban areas is possible, e.g., some recent research was realized under the assumption of exploiting shadowing effects created in the communication environment (e.g., by buildings, obstacles) in favour of frequency reuse \cite{Bedogni2014}.

Since different standards for opportunistic communication in the TV White Spaces have now been published, such as IEEE 802.22 \cite{Stevenson2009}, IEEE 802.15.4m \cite{Funada2012}, Weightless \cite{Weightless2012}, and of course IEEE 802.11af, improved methods to guarantee the coexistence of different devices, operating on several protocols in the same bands, must be deployed. We highlight here some work, e.g., \cite{Ghosh2011} \cite{Kulac2009} \cite{Villardi2011} \cite{Wang2012}, \cite{Bedogni2014}, that focuses on the coexistence between different technologies operating in the same bands. In particular, \cite{Kulac2009} studies the performance degradation of IEEE 802.22 when an IEEE 802.11af network is operated in the same area. The problem becomes even worse when the IEEE 802.11af network is located near 802.22 user equipment, causing both networks to perceive a strong interference from each other. One proposed solution is the Coexistence Beacon Protocol \cite{Henderson2009}, studied for 802.22 networks, which foresees the exchange of a periodic beacon to identify the neighbouring, and possibly interfering, networks. 

For what concerns the co-existence problems, research has addressed other important related aspects of new WLANs on TV White Spaces. For instance, the higher coverage range makes them attractive for applications in the smart grid, and also for M2M on TVWS \cite{Bedogni2014a} \cite{Bedogni2014} \cite{Liu2012}. Here, the increased range compared to standard technologies in the ISM bands can have beneficial effects for indoor mobile devices \cite{Bedogni2014}, and M2M \cite{Liu2012}. Specifically, indoor devices communicating on TV White Spaces offer better propagation characteristics and penetration through obstacles, making it easier to realise scalable home connectivity \cite{Novlan2010}, although interference and co-existence problems could be exacerbated and must be resolved, e.g. see \cite{Simic2011}. However, studies in the literature show how the signal coming from an indoor transmitter on TV White Spaces remains constrained inside the house, making it difficult to deploy indoor-to-outdoor networks. On the other hand this shadowing limit can be seen as an opportunity, lowering the interference outside the building in which the TVWS network is locally operated \cite{Bedogni2014a} \cite{Bedogni2014}. Under this approach, HDTV streaming has recently been considered as a possible use case of WLAN networks operating in the TV White Space \cite{Fadda2012}.
%
%
\paragraph{Non-contiguous channel bonding}
\label{sec:ncofdm}
A relevant novel feature of IEEE 802.11af is the potential for contiguous and non contiguous channel bonding, which permits aggregating basic channels (also non adjacent ones), and leveraging the possible large frequency spread between multiple available channels. Due to the rather static nature of primary DTV transmissions, where a busy channel is unlikely to become free in the near future, and state changes are coarse-grained in general, it is crucial to exploit the time-locality effect and exploit the maximum physical channel availability that could be aggregated at any given location.  With the methodology inherited from IEEE 802.11ac, IEEE 802.11af is capable of bonding together two up to four basic channels grouped in up to two different non-contiguous chunks \cite{Flores2013}. The spectrum bandwidth of a DVB-T basic channel can be either 6, 7, or 8 MHz, depending on the country in which the service is operated. As an example, this creates a 144 to 168 OFDM channels' bandwidth potential when up to four 6-7-8 MhZ channels are bonded \cite{Flores2013}.
%
%
%
\subsubsection{Open Challenges} \label{sec:openchallenges}
\noindent
The deployment of WLANs in the TVWS at their maximum potential still requires the resolution of open problems. Generally speaking, as mentioned in previous sections, the challenges for such networks can be divided into three main categories: 1) spectrum sharing between opportunistic (secondary) and heterogeneous devices, 2) spectrum sensing/management and maximum exploitation of available spectrum (potentially enabling the spectrum-on-demand concept), and 3) co-existence and interference mitigation to the primary network (primary user protection) and secondary networks \cite{Ghosh2011}.

The overall aim of spectrum sharing techniques is to maximise separation to avoid overlapping or contiguous channels operations, provided that sufficient channels are available in the TVWS spectrum in a given space/time scenario \cite{Ghosh2011}. In general, important research contributions have to be realised for the \emph{design of proper spectrum allocation methodologies and techniques} satisfying multi-factorial QoS requirements at the system-level and the user-level. Novel ideas include the spectrum sharing and spectrum sensing techniques being further divided into cooperative and non-cooperative. 

Non-cooperative solutions attempt to realise spectrum sharing and sensing on a local basis, without direct cooperation between devices. Examples include the "listen-before-transmit" approach, and transmission power control to limit the interference, and spectrum allocation policies that are based on local node feedback only.  Cooperative solutions rely on the existence of a common communication channel, through which devices can tentatively agree on the spectrum allocation that provides the desired spectrum separation and QoS. Major issues include the definition of a methodology to identify a common channel (or multiple channels mutually shared in space by pairs of heterogeneous and distributed devices) and to efficiently share common information. For the access to the spectrum DBs, the identification of proper common access channels is still under discussion: out-of-TVWS-band cellular communications can be widely used, given their high coverage, also indoor. 

Trying to avoid the use of complex and resource-hungry coordination schemes, and dedicated control channels, \emph{rendezvous protocols} play a significant role, both for co-existence and for establishing dynamic communication channels, based on spectrum sharing. Rendezvous is considered a fundamental problem in generalised CR networks, at the basis of many communication processes, e.g. including neighbour discovery, routing and broadcast. Rendezvous protocols attempt to establish a new link for communication on a selected and agreed frequency band (channel) identified within a set of available resources (channels), thus creating the basis for communication between two or more SUs. This problem is often exacerbated by the high (and unknown) numbers of contending SUs, high number of available channels and their ambiguous boundaries and identifiers, and tough co-existence requirements satisfaction in cognitive radio systems \cite{Hoi07}. Preliminary approaches for rendezvous were based on the existence of a dedicated Common Control Channel (CCC) agreed among all the SUs, or the existence of a centralised agent (decision maker) assigning channels to any pairs of requesting users. This approach could be considered in the case of IEEE 802.11af solutions based on the spectrum DB implementation, however, the problem is complicated by the fact that the common channel must be identified and made available for all the requesting SUs at the same time, even when these are spread over multiple collision domains in space. On the other hand, the scalability and reliability issues which are caused by the congestion of CCCs, and the vulnerability risks of centralised DBs, motivate the research of alternative distributed solutions, called blind rendezvous protocols \cite{Gu2014}. Distributed solutions have been mainly based on beaconing and/or slot-based channel hopping algorithms exploiting randomisation, sensing and discovery/synchronisation messages spread in slotted time over a set of candidate available channels \cite{Hoi07}. These protocols aim to minimise the convergence time, to maximise the use of resources, and to avoid collisions among SUs, by identifying and agreeing on the use of a given common channel among N non-overlapping and unambiguously labeled channels. The main complication in IEEE 802.11af systems is given by the wide availability of many different non-contiguous spectrum bands, which cannot be uniformly quantised in single channels and unambiguously labeled in space, in order to realise a common reference domain for all the IEEE 802.11af devices. In fact, the knowledge of the number of channels, their common labels, the unique IDs of SUs and the number of SUs contending for the channel assignment are the key factors determining the effectiveness and good properties of blind rendezvous protocols. Recently, new classes of rendezvous protocols have been proposed which are distributed, blind and oblivious. Oblivious means that available channels may be identified with different labels on behalf of the different involved SUs. A number of these distributed blind and oblivious rendezvous protocols have been proposed and have been referenced and analysed in \cite{Gu2014}. To conclude, solutions for dynamic spectrum handoff issues (e.g. in case PUs suddenly appear in a region) must be resolved in advance to minimise the latency of re-establishing a functional communication channel, e.g. by implementing a proactive alternative rendezvous channel selection in background to ongoing communication processes \cite{Ren2012}.

Another challenge is to provide \emph{cooperative techniques in a distributed vs. centralised implementation}, by analysing their convergence under variable conditions, overheads and tradeoffs. In some cases, as in vehicular networks, the exploitation of scenario characteristics and factors such as the constrained mobility could help to realise effective dissemination of sensing information over a extended and predictable horizon under a cooperative approach \cite{DiFelice2013VTC}, \cite{DiFeliceVNC2011}, \cite{DiFelice2010WD}. Cooperation could be helpful in the realisation of the cooperative spectrum sensing in combination with the Geolocation DB approach. Another interesting direction of research is the investigation of mutual effects of coexisting cooperation-based vs. non-cooperative techniques. In cooperation-based devices the exchange of information could allow a quick convergence to a solution, and avoidance of further interference \cite{Chowdhury2011}. Certainly, cooperative techniques have desirable advantages over non-cooperative ones. However, tight time synchronisation and high coordination overheads are required. The common channel for cooperation purposes could become a bottleneck and reduce the potential advantages of dynamic spectrum allocation. The IEEE 802.19 foresees the presence of a shared common control channel (CCC), to which devices need to tune in order to gather the state of the network. Solutions such as \cite{Bedogni2013} propose the use of TVWS for the CCC. 

A promising direction for research is the adoption of clustering schemes for secondary devices, associated with a hybrid distributed/cooperative vs. centralised/non-cooperative approach, in an effort to reduce overheads while maximising advantages. In this way, secondary nodes could implement cooperation for sensing functions using properly identified common channels, and delegate the centralised decisions and spectrum-DB management to well instrumented cluster-leader nodes. This concept is incarnated to some extent by RLSS in IEEE 802.11af. More recently there have been proposals to build distributed databases that refresh their contents periodically by querying the remote spectrum database \cite{Bedogni2014b}. Here, Master devices (according to Ofcom terminology) periodically cache the query replies from the spectrum database in order to reply to and broadcast the spectrum availability to neighbour Slave devices. 

Another challenging direction for research is the definition of \emph{accurate models and efficient simulation tools enabling dynamic spectrum analysis for both rural and urban areas}. Recent developments of digital maps, simulation tools and propagation models theoretically allow improving the capability of predicting radio propagation effects in complex scenarios with an acceptable computation time. This enabling technology could be used in parallel to spectrum DBs to provide more accurate forecasting of dynamic frequency allocation in time/space scenarios. Another issue to be taken into account is the accuracy of the remote spectrum database regarding the channel availability estimation. Several works have already shown the inaccuracy of propagation models, due to the complexity of parameters to be considered and differences between modelled and real scenarios~\cite{McHenry2006,Avez2012,Bedogni2014a,VandeBeek2012}. Better propagation models will make it possible to more efficiently estimate the interference between devices, and build Radio Environment Maps (REM) or White Space Maps (WSM) that will possibly lead to a more accurate sharing of the radio spectrum \cite{Akyildiz2011}. To this end, current research could focus on how to leverage simulation models in order to build more accurate frequency DBs.

In general, the adoption of remote spectrum DBs methodology pushed by national regulators has made the spectrum sensing process optional and conservatively realised (e.g., power control for sensing-only devices is limited to below 17 dBm EIRP). Thus, much of the research has focused on techniques and \emph{solutions to successfully query the remote database}, also in challenging environments, such as rural areas and indoor. However, since the queries must provide the position of the mobile device, the accuracy of positioning for mobile and handheld devices still needs more appropriate solutions. Triangulation can be used to estimate the position of mobile devices with the accuracy needed by the current regulations outdoor (50m). However, a low GPS accuracy is possible indoors \cite{Zandbergen2009}. Certified manual positioning can be provided by technicians for static and indoor devices, although limiting mobility. Alternatives to costly infrastructures for positioning (e.g., based on short range beaconing devices) must be deployed in indoor scenarios.

Finally, regarding \emph{spectrum sensing and interference mitigation}, new challenges come from the heterogeneity and different characteristics that networks operating in the TV White Spaces can have, such as different transmitting power (4 W for static devices, 100 mW for mobile and portable devices, and 40 mW for communication in channels adjacent to occupied ones), different bandwidths (5, 10, 15, and 20 MHz are currently foreseen for IEEE 802.11af; 802.15.4m can have narrow bandwidths or wide ones), and different medium access schemes (protocols that use either CSMA or TDMA are required to co-exist in the TV White Spaces). With all this in place, the problem of determining whether another secondary device is currently transmitting in the same channel is still an open issue.

Spectrum sensing is a very hot topic in the literature, with many proposals that leverage the cooperation between devices to raise the accuracy of the detection process, e.g., \cite{Akyildiz2011, DiFelice2010a}. The current literature also offers room for proposals that suggest exploiting sensing to build more accurate spectrum DBs \cite{DiFelice2013VTC} \cite{Doost-Mohammady2012} \cite{Gurney2008a}. Lastly, regarding the interference mitigation to the primary network, regulators provide a conservative approach based on a remote spectrum database to avoid high risk for the primary users. However, in particular for the indoor scenario, the concept of secondary users' transmission in TV Grey Spaces has been proposed.  TV Grey Space identifies busy channels that are formally used by primary users in the area of interest (e.g., at the rooftop level of a building for DVB-T), but which could be considered usable indoors (e.g., at the basement level), without causing effective interference to the primary users \cite{Bedogni2014a,Bedogni2014,Peha2013}. This concept is based on the generalised spectrum underlay paradigm \cite{LBLe2008}. The remaining research challenge is to provide TV Grey space access guaranteeing a sufficient amount of protection to the primary user, while providing a satisfactory QoS to opportunistic devices, e.g., through the use of customised and validated propagation models for indoor TVWS networks, and feedback mechanisms to constantly monitor the interference generated to the primary network.
%
%
%
\section{Emerging new trends and technologies} \label{Sec:Emerging}
\noindent
In this last section we review three emerging trends that, in our opinion, will have a large influence in the conception of future WLANs as they change the way WLAN protocols and functionalities are developed, implemented, tested and integrated with other wireless networks.
%
%
\subsection{Programmable Wireless LANs}
\noindent
Especially in the enterprise environment, WLAN deployments need to support a wide range of functionalities and services. This is intrinsically difficult because of the large number of APs that must be managed, which calls for scalable solutions. Typical services include channel assignment, load balancing among APs, authentication, authorisation and accounting (AAA), policy management, support for client mobility and interference coordination. Another problem is the fact that WLAN clients autonomously take several decisions such as which APs to associate with, when to hand-over, etc. Therefore, supporting roaming clients requires the management of a large number of association states across several APs, which is a challenge if support for real-time hand-over is desired. Typically, such management schemes are centralised and most of them are proprietary, such as WLAN controller solutions from Aruba~\cite{ARUBA} or Cisco~\cite{CISCO}, although the 802.11u amendment has been released to allow mobile users to seamless roam between WiFi networks with automatic authentication and handoff~\cite{2014-comcom-80211u}. For example, Dyson \cite{Murty:2010:DAE:1855840.1855855} enables STAs to send information such as radio channel conditions to a centralised controller based on a custom API (e.g., based on Python). As the controller has a centralised view of the network, it can enforce a rich set of policies to control the network also using historical information. A demo system has been implemented along with applications such as airtime reservations for specific clients or optimised handoffs. However, Dyson requires STAs to be modified in order to use those new services offered by the centralised controller. TRANTOR~\cite{TRANTOR} is another example of a centralised management system that requires changes of clients in order for the infrastructure to gather information from them in terms of e.g., interference measurements. In addition, control commands enable the infrastructure to exercise control such as modifying the transmit power or influencing the association procedure. Clients still use standard CSMA MAC layer for data transmission. In contrast to DenseAP~\cite{DenseAP} or DIRAC~\cite{Zerfos:2003:DSW:938985.939009}, which are also based on a centralised controller but do not modify clients, TRANTOR exercises much more control on clients by the use of a dedicated API, which enables a significantly higher gain in coordination and thus capacity. CENTAUR~\cite{Shrivastava:2009:CRF:1614320.1614353} is another example of a centralised controller that centrally schedules hidden and exposed terminals in the downlink while using standard CSMA MAC for uplink traffic and legacy downlink traffic. To achieve good performance, it uses a fixed back-off, packet staggering techniques and a hybrid data path that only schedules downlink transmissions towards hidden and exposed terminals centrally. All other traffic is sent using standard DCF with a standard back-off procedure. In contrast to previous approaches, it does not require any modifications to the clients but does require data plane centralisation and it remains questionable how scalable such a solution is for large WLAN deployments with high PHY layer data rates beyond 100 Mbps. However, new approaches for fast data plane processing that are currently explored in the virtualisation community which move all the packet processing into user space may be an option to significantly increase speed. Designing such management systems, raises several interesting research questions: (a) Centralized infrastructures may incur high latency but are more simple to program and maintain while designing a completely distributed approach that operates close to WLAN devices may result in significant distributed coordination and consistency problems; (b) what level of distributed control is required in order to support the flexibility needs of future WiFi based networks supporting a high level of mobility.

Traditionally, WLAN APs are built on proprietary operating systems that are tightly coupled with the hardware. This design makes it hard to create new applications on top of such networking devices. Despite the fact that protocols and mechanisms are available that would greatly improve the utility of existing networks, those new protocols are not deployed, because the closed system design makes it very difficult to extend their functionality. An important aspect for all such new approaches is therefore how to provide open interfaces and open source to speed up innovation. As an example, Linux-based devices are completely open source but in order to increase flexibility, much more work is needed in the area of open drivers and firmware. In fact the Atheros based ath9k driver has been the main driver of innovation in the open source community because it's the only driver that interworks with open firmware. \textcolor{black}{In general solutions that modify the OS driver focus mainly on programmable network level solutions for WLANs, such as channel switching or handovers between APs. While the ath9k driver is an excellent example in how to speed up innovation, it is still very cumbersome to support flexible MAC engine reprogrammability with ath9k. In contrast, MAClets and Wireless MAC processors (WMP)~\cite{tinnirello2012wireless} allow a much more flexible reprogramming of MAC functionalities. A MAC processor is an entity able to execute general MAC commands that specify the MAC operations through a software-defined state machine. The behaviour of the MAC protocol can therefore be updated at run-time by simply changing the sequence in which those commands are executed (i.e., the MAClets). As a proof-of-concept, the authors implement the proposed MAC processor solution in a commodity WLAN hardware card, extending the basic DCF in three directions: piggy-backing ACKs, a pseudo TDMA, and the use of multiple channels. In~\cite{2014-conext-maclets} a control framework for this WMP system is also proposed to support MAClet code mobility, i.e., for moving, loading and activating MAC software programs embedded into ordinary data packets (akin to the capsule model of traditional active networks).} 

The difficulty in re-programming networking hardware has also led to the concept of Software Defined Networks (SDNs) based on the OpenFlow protocol~\cite{OpenFlow}. The main idea of SDNs is to extend networking devices with standardised APIs that allow third-party programmers to flexibly control the data path. In addition, SDNs provide higher level abstractions to network designers and programmers through the use of a centralised control plane offered by a network controller such as NOX~\cite{NOX}, which allows reuse of components such as topology discovery or network access control for different applications. The SDN concept has recently been applied to WLAN architecture in order to enable fine grained control over mobility management and data forwarding focusing on programmable enterprise WLAN architecture. An important part of the SDN architecture is the network controller, which provides a centralised view of (parts of) the SDN enabled network and uses the OpenFlow protocol to install the forwarding rules on SDN-enabled devices (routers, switches, access points, cellular base stations, etc.).  

\begin{figure}[t!!!!!!!!]
\centering
\includegraphics[angle=0,trim=0cm 0cm 0cm 0cm,clip=true,width=5cm]{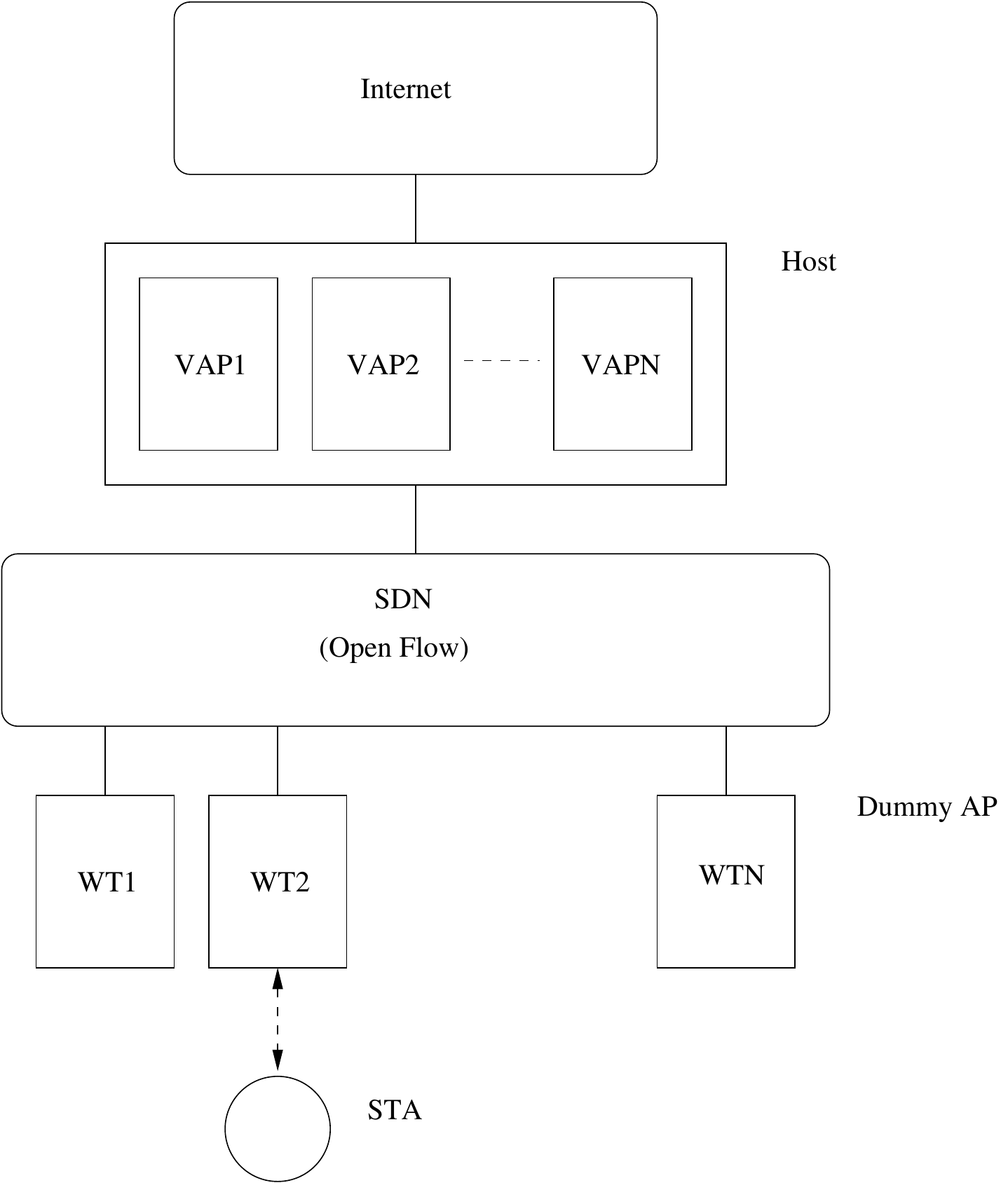}
\caption{The CLOUDMAC architecture. VAP stands for Virtual Access Point, and WT stands for Wireless Termination.}
\label{Fig:CloudMAC}
\end{figure}

ODIN~\cite{ODIN} is designed to support programmability in enterprise WLAN architecture by separating the association state from the physical AP. They implement Light Virtual Access Points (LVAP) using a Split-MAC approach, where the infrastructure controls the handover procedure among different WLAN APs. By managing associations through SDN controllers, ODIN enables proactive mobility management and load balancing within the SDN enabled WLAN enterprise network without the need for changes in the client WLAN stack or IEEE 802.11 MAC layer. While ODIN requires agents to reside on the APs that communicate with the Odin Master within the SDN controller, CLOUDMAC~\cite{CLOUDMAC} is based purely on the concept of SDN and virtual APs. Similar to ODIN, CLOUDMAC implements a Split-MAC approach but in addition enables the processing of WLAN MAC layer frames within a co-located Cloud using so-called Virtual APs (VAPs). The physical APs in CLOUDMAC are lightweight WLAN APs that are responsible only for sending out IEEE 802.11 based MAC layer ACKs to standard WLAN clients and tunnel WLAN MAC layer frames through an SDN-enabled enterprise WLAN towards the VAPs. As association states are kept in the VAPs, fast mobility is supported using simple OpenFlow forwarding rules. Because of the additional processing of IEEE 802.11 WLAN MAC frames in the co-located Cloud, CLOUDMAC has higher latency than a standard WLAN deployment. However, CLOUDMAC offers a Webservice based API to third party applications in order to program the enterprise WLAN and enable new services. The architecture of CLOUDMAC is depicted in Figure \ref{Fig:CloudMAC} and has been extended in~\cite{7116164} to support QoS management and in~\cite{CLOUDMACPRIO} to support flexible MAC management frame prioritisation based on 802.11. A system based on OpenFlow has been recently proposed in~\cite{2014-comcom-bestap} to allow the station to be associated with multiple APs simultaneously and to switch between APs with low overhead.

An important aspect to consider in order to enable programmability is backwards compatibility. Several approaches (e.g.,~\cite{TRANTOR}) require clients to be modified for them to utilise the features provided. This is difficult to do in practice because it requires changes in the operating system software of all clients. If not all clients can utilise those APIs, it is questionable how usable the new architecture will be and how much benefit in terms of aggregate performance such architecture will allow. In contrast, the SDN based approaches are interesting in the sense that they do not require changes in the client WLAN stack and work with the standard IEEE 802.11 MAC layer deployed within the clients. However, it remains questionable how scalable such solutions are. For example, an interesting open research topic is to evaluate the scalability of approaches such as CLOUDMAC~\cite{CLOUDMAC}. In addition, processing IEEE 802.11 MAC frames within co-located private Cloud requires low latency support from local Cloud solutions (such as OpenStack) in order to reduce the IEEE 802.11 MAC processing time, which is an area of active research.  
%
%
%
%
\subsection{Prototyping and testing IEEE 802.11 enhancements}
\noindent
Most of the new proposals for next-generation WLANs are currently only evaluated using mathematical analysis and simulation. While both analysis and simulation are necessary to characterise and study those enhancements in the initial design phase or to consider large-scale scenarios, it is difficult to consider all practical aspects of a real-world scenario. This can sometimes cause significant differences between what simulations and real experiments show. However, real experiments are challenging because of the high complexity and costs of building the new hardware and software for each specific solution to test.

To mitigate implementation complexity, there are several flexible hardware platforms where low-level MAC mechanisms can be completely implemented in software. Examples based on FPGAs are the USRP (Universal Software Radio Peripheral) \cite{USRP} and the WARP (Wireless Open-Access Research Platform) \cite{WARP}. As and alternative, SORA (Microsoft Research Software Radio) \cite{SORA} works on general purpose computers by taking some advantage from modern multiple core systems. 

A different approach is provided by OpenFWWF \cite{OpenFWWF}. OpenFWWF provides an open CSMA/CA firmware for specific models of Broadcom chipsets, so the resulting firmware can be uploaded and tested in real commercial hardware. OpenFWWF implements a simple State Machine (SM) for controlling the hardware in real time. The SM evolution is driven by a main loop that reacts to events by executing specific handlers. When a packet, originally prepared by the Linux kernel, is ready in the NIC  memory, handler \textsc{Packet$\_$Ready} sets up the radio hardware according to the packet metadata (e.g., it fixes rate, modulation format, and power level), schedules the transmission and jumps back to the main loop. The Transmission Engine (TXE) then takes care of accessing the channel, i.e., it decrements the back-off counter according to the DCF rules and eventually starts the actual transmission. This triggers the execution of the \textsc{TX$\_$frame$\_$now} event that prepares the ACK time-out clock and finalises the MAC header (as the transmission has already started, these actions must be completed before the end of the physical preamble). If the ACK-frame is received or if the ACK time-out expires and the maximum number of attempts for this packet is reached, handler \textsc{Update$\_$Params} resets the contention window to its minimum value, or otherwise doubles it. Finally, it loads the back-off counter with a fresh value. In the following, we will review a list of selected papers that use the OpenFWWF firmware and the WARP or USRP platforms to test new proposals for WLANs in real scenarios. 

A first implementation of the groupcast mechanisms defined in IEEE 802.11aa is presented in \cite{salvador2013first}. The authors modify the OpenFWWF firmware to include those functionalities and evaluate them using a 30 STAs testbed. A second example of the use of OpenFWWF is the collision-free MAC protocol presented in \cite{sanabria2014implementation}. In \cite{sanabria2014implementation}, the authors design and implement in OpenFWWF a MAC protocol that is able to achieve a collision-free operation by waiting for a deterministic timer after successful transmissions. This work shows the experimental results of a collision-free MAC (CF-MAC) protocol for WLANs using commercial hardware. Testbed results show that the proposed CF-MAC protocol leads to a better distribution of the available bandwidth among users, a higher throughput and lower losses than the unmodified WLANs clients using a legacy firmware. 

In \cite{magistretti2012802}, the authors use the WARP platform to test a variant of the DCF -- called IEEE 802.11ec -- that substitutes control packets such as the RTS, CTS and ACK with short detectable sequences. Since those sequences are shorter than the control packets, and can be detected correctly even at lower SNRs values, a significant gain in performance is achieved. MIDAS (Multiple-Input Distributed Antenna Systems) \cite{xiong2014midas} implementation using the WARP platform shows the benefits of distributing the antennas over the area to cover instead of co-locating them at the AP in terms of capacity when MU-MIMO is employed. The authors also propose a new MAC protocol to benefit from the spatial reuse that the DAS allows, taking as a basis the IEEE 802.11ac. Results in a testbed show that their proposal is able to achieve up to 200 \% gains versus the traditional approach where all antennas are co-located at the AP.

The USRP platform has recently been proposed to develop a first implementation of IEEE 802.11ah in order to obtain a preliminary performance assessment of such a technology, because no commercial off-the-shelf hardware is yet available as the amendment is still in progress \cite{aust2014advances}. The USRP platform is also used in \cite{gollakota2011clearing} to evaluate TIMO (Technology Independent Multi-Output) a solution to deal with high-power non-IEEE 802.11 interferers in ISM bands. 

\textcolor{black}{To conclude this section, it is worth mentioning here also to MAClets and WMP, as given such flexibility to implement and distribute new MAC protocols and other IEEE 802.11 functionalities, we believe that if a use-friendly implementation of such a MAC processors framework is provided, it would significantly contribute to the development and testing of new MAC enhancements by the research community.}
%
%
%
%
%
\subsection{Cellular/WLAN interworking}
\begin{figure}[t!!!!!!!!]
\centering

\includegraphics[angle=0,trim=0cm 0cm 0cm 0cm,clip=true,width=7cm]{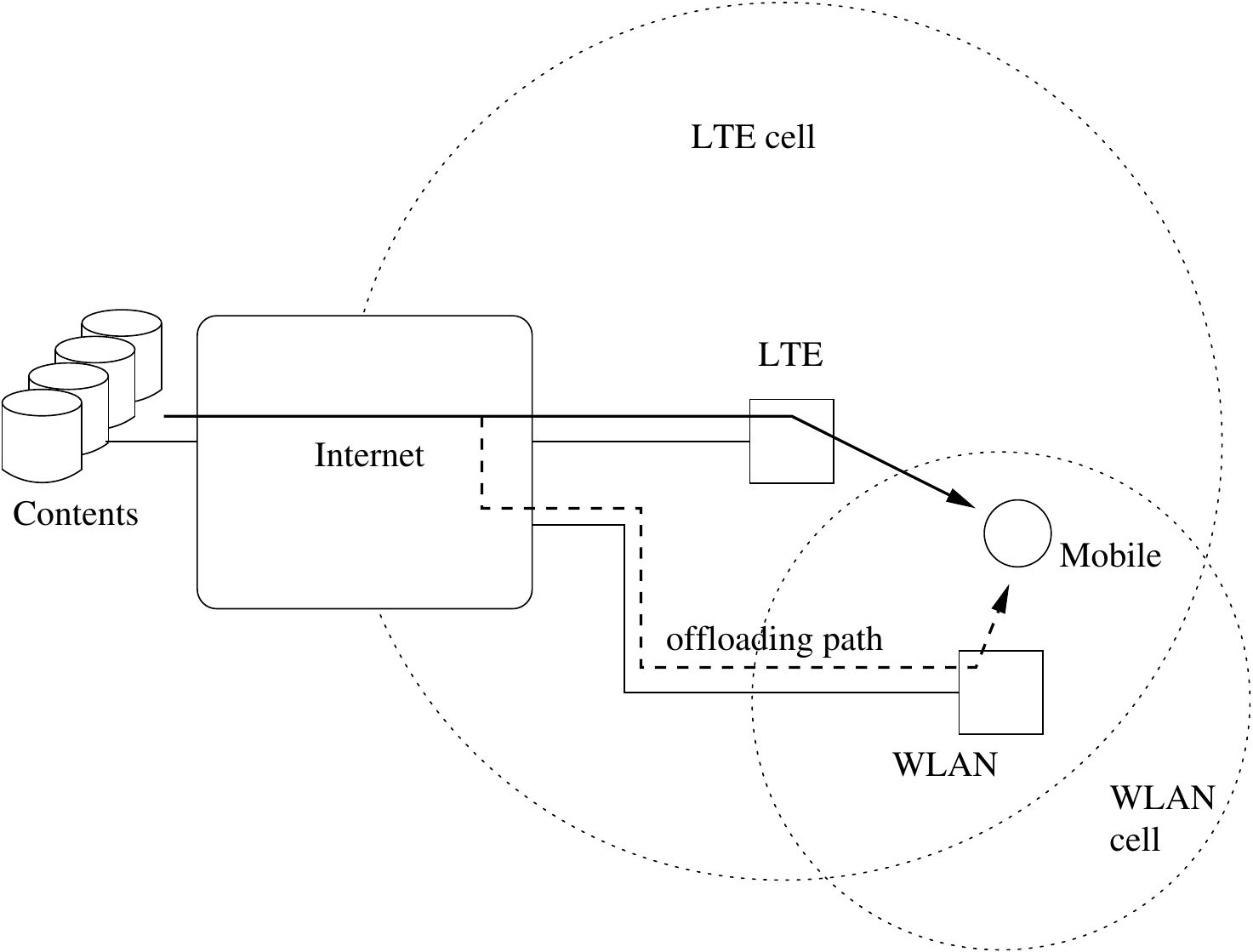}

\caption{LTE to WIFI offloading.}
\label{Fig:LTEWIFIOff}
\end{figure}

\noindent
Public hotspots that offer Internet access over a WLAN using IEEE 802.11 technology are now nearly ubiquitous. It is forecasted that the cumulative installed base of WiFi hotspots worldwide will amount to 55.1 millions by 2018,  excluding private hotspots (e.g., WiFi access points deployed at home)~\cite{2013-techrep-wba}. The sharp increase in the availability of public WiFi was initially perceived by mobile cellular operators as a threat due to the additional competition from wireline Internet service providers or emerging crowdsourced WiFi networks, such as FON\footnote{https://corp.fon.com/en.}. However, as cellular operators are fighting to cope with the explosion of mobile data traffic created by the rising use of multimedia content traffic over mobile devices~\cite{2013-cisco-report}, they are also starting to use WLANs based on the IEEE 802.11 technology to \emph{offload} data from their core and access networks. In general, mobile data offloading refers to the use of complementary network technologies (in licensed or unlicensed spectrum) for delivering data originally targeted to cellular networks. Intuitively, the simplest type of offloading consists of exploiting connectivity to existing co-located WiFi networks and transferring data without any delay (Figure \ref{Fig:LTEWIFIOff}). Thus, this offloading technique is know as \emph{on-the-spot offloading}. As a consequence of this new trend, the seamless integration of cellular (e.g., 3G/LTE) and WiFi technologies has attracted significant research interest in recent years (see~\cite{2014-cst-survey-offloading} for a survey), a few solutions have already been standardised~\cite{2012-cst-LIPA-SIPTO,AstelyDFPS13}, and roaming between cellular and WiFi is becoming increasingly transparent to end users. Cellular/WLAN interworking is also fostered by the support in the evolving 4G standards of heterogeneous network deployments (HetNets), in which the existing macro cells are complemented with a number of small, low-power base stations with the goal of increasing capacity in highly congested areas~\cite{2011-review-hetnets,2013-commag-hetnets}. It is envisaged that small cells will be based on 4G standards (e.g., pico and femto cells) as well as IEEE 802.11 technologies, and multimode base stations that work simultaneously with LTE and WiFi are already entering the market.

It is important to point out that LTE standards already support a variety of mechanisms that enable data offloading. However, most of the existing solutions, such as Local IP Access (LIPA) and Selected Internet IP Traffic Offload (SIPTO), focus on data offloading at the core of the network~\cite{2012-cst-LIPA-SIPTO}. For instance, LIPA allows an IP-enabled mobile device to transfer data to another device in the same pico or femto cell without passing through the cellular access network, while SIPTO enables the routing of selected IP data flows through different gateways. The only LTE offloading mechanism that supports seamless interworking with IEEE 802.11-based WLANs is IP Flow Mobility and Seamless Offload (IFOM). Specifically, IFOM relies on Mobile IPv6 technologies to allow a user terminal to simultaneously route selected IP flows over different radio access technologies~\cite{3GPP23.261,2011-commag-ifom}. In this way, a user terminal can offload selected flows to a WLAN based on some operator-defined policy while keeping the LTE connection running. However, to enable such an approach it is necessary to support an entity in the cellular core network that can communicate with the user terminal to exchange information about the availability and quality of neighbouring access networks, as well as to provide the user terminal with predefined rules to manage the handover process. This entity in current LTE standards is called the Access Network Discovery and Selection Function (ANDSF) server~\cite{3GPP24.302}. Note that the problem of selecting the best communication technology in an heterogeneous wireless network has been extensively investigated in the literature both with centralised as well as decentralised approaches (see~\cite{2013-cst-network-selection} for a survey). However, the use of multiple interfaces in parallel, as well as per-flow offloading are relatively new concepts. Thus, the design of scalable and efficient network selection strategies for the ANDSF framework is still an open issue. For instance, three offloading methods suitable for the ANDSF framework are proposed in~\cite{2013-wmnc-andsf}, which are based on coverage, SNR, and system load. A reinforcement learning approach is designed in~\cite{BennisSCSVD13} that allows multimode base stations to autonomously steer their traffic flows across different access technologies depending on the traffic type, the users' QoS requirements, the network load, and the interference levels. A number of studies have also tried to quantify the potential capacity gain of WiFi offloading in real-world WiFi deployments. For instance, the authors in~\cite{2013-ton-wifi-offloading-lee} evaluate offloading efficiency using trace-based urban mobility patterns and WiFi connectivity distributions. The authors in~\cite{2012-vtc-offloading} instead analyse the offloading performance in an indoor scenario in which femto cells and WiFi access points coexist. On the other hand, mathematical models are needed to derive performance bounds and guide the design of optimal offloading strategies. A simple queuing model for the analysis of the offloading efficiency is developed in~\cite{2013-globecom-model-offloading} by assuming that the WiFi network availability is exponentially distributed. A more general scenario is considered in~\cite{2013-twc-model} by assuming multiple classes of access points that differ in transmit power, deployment density and bandwidth. The optimal association strategy is then derived to maximise the fraction of time that a typical user in the network is served with a rate greater than its minimum rate requirement. A somehow related problem consists of deciding how to optimally deploy WiFi hotspots to maximise offloading efficiency. Traditionally, this problem has been considered from the point of view of coverage maximisation, e.g., to ensure continuous WiFi connectivity by taking into consideration user mobility characteristics~\cite{2013-tvt-ap-deployment}. On the contrary, in the context of mobile data offloading, AP deployment is tackled to maximise the throughput performance in an heterogeneous wireless network. For instance, a heuristic algorithm is proposed in~\cite{2012-pingen-ap-deployment} that selects as AP locations the cells with the higher frequency of download requests. A graph-theoretical solution for AP deployment is developed in~\cite{2014-secon-ap-deployment} by consorting a time-dependent graph that describes the interdependencies between users' mobility trajectories, points of interest and traffic demands. An open issue in this field of research concerns the design of more adaptive traffic steering mechanisms between cellular and WiFi. Furthermore, the increase in the number of wireless infrastructure nodes with the dense deployment of small cells will make the future network deployments quasi stochastic. Thus, new methodologies, such as stochastic geometry, have to be explored to model the performance bounds of heterogenous wireless networks that allow the inter-working between cellular systems and WLANs~\cite{2013-cst-stochastic-geometry}. Finally, the use of historical data to predict network conditions and user locations may also become infeasible due to the scale of the network in terms of infrastructure nodes and users. To deal with this issue limited measurements could be coupled with statistical inference methods.

On-the-spot offloading is the dominant but not the only form of interworking between cellular networks and WLANs. More recently, \emph{delayed offloading} has also been proposed for delay-tolerant traffic. Basically, if a user is willing to accept a delayed content reception (e.g., the download of a YouTube video), the cellular operator may intentionally postpone the content transfer in order to wait for WLAN availability or better transmission conditions. The cellular network is then used to complete the data transfer only if the content reception can not be guaranteed within a user-specified deadline. A number of studies have explored the feasibility of delayed offloading for different delay deadlines using trace-based WLAN usage patterns. In particular, results in~\cite{2013-ton-wifi-offloading-lee} confirm that increasing the delay-tolerance of content significantly improves the fraction of traffic that can be offloaded. The offloading performance clearly depends on several factors, including the location of  802.11 hotspots and the ability to accurately predict the future availability of WLAN coverage. Thus, several studies have addressed the problem of forecasting mobile connectivity. For instance, the solution proposed in~\cite{2008-mobicom-breadcrumbs}, called \textsc{BreadCrumbs}, tracks the movements of the mobile device's owner and maintains a history of observed networking conditions to train a forecast model of near-term connectivity. More recently, a time-based prediction model of visited locations derived from movement traces of mobile users is given in~\cite{2013-ton-mobility}. Then, the authors in \cite{2010-mobisys-wiffler} propose a system, called Wiffler, that allows mobile users to decide whether or not to wait for a future WiFi offloaded opportunity based on the predicted WLAN capacity and the total data that needs to be transferred. However, the design of location prediction models that have low computation complexity and are suitable for short-term mobility is an open issue. Also related to the problem of delayed offloading feasibility is the optimal placement of APs. The authors of~\cite{2011-mass-ap-deployment} develop the \textsc{HotZones} algorithm, which selects the cells with the highest number of daily visits as the location of additional WiFi access points. A similar solution, called \textsc{Drop Zones}, is proposed in~\cite{2012-tmc-dropzones}. In the context of vehicular networks, the optimal deployment of RoadSide Units (RSU) over a given road layout to maximise the overall system throughput is analysed in~\cite{2013-tmc-fiore-vehicular-downloading}. In~\cite{2015-comcom-offloading-vehicular} it is analysed the data offloading gain in a vehicular sensor network as a function of the percentage of equipped vehicles, of the number of deployed road side units, and of the adopted routing protocol. Although existing work established the potential of delayed offloading, there is still a need for considerable research to design incentive mechanisms for motivating users to leverage their delay tolerance for cellular traffic offloading. For instance, an auction-based pricing framework is proposed in~\cite{2014-tmc-offloading-incentive} to give priority to users with high delay tolerance and a large offloading potential. 

We conclude this section by discussing a third type of mobile data offloading, known as \emph{opportunistic offloading}, which does not rely on a WLAN infrastructure but exploits direct communications between mobile devices, e.g., through the emerging WiFi Direct standard~\cite{2012-wc-d2d-wifi-direct}. Opportunistic offloading schemes allow saving significant cellular bandwidth because the content spreads through the opportunistic network formed by the users, while the cellular network is mainly used for signalling and triggering the content dissemination. Clearly, the performance of opportunistic offloading solutions depends on several factors, including the user mobility patterns, the user density, the delay tolerance for content reception, and the popularity of the content that needs to be transferred. Note that opportunistic offloading requires a less controlled type of interworking between LTE and WLANs because user devices can setup direct connections and start ad hoc communication autonomously with little or no intervention from the operator. For instance, the authors of~\cite{2011-mass-ap-deployment} propose a simple algorithm, called \textsc{MixZones}, to allow the cellular operator to decide when the mobile users should be notified to switch their wireless interface for data transfer with potential other users that they are predicted to encounter. In addition, most of the papers consider that content must be delivered to users within a given deadline. Most of the research in this context addresses the problem of selecting the best (e.g., the smallest) set of users, called seeds, that should receive the content from the cellular network and help to disseminate it over the opportunistic network. Three simple algorithms for initial seed selections are proposed in~\cite{2012-tmc-offloading-han}. The authors of~\cite{2014-adhoc-marcelo-offloading} propose using social network properties, e.g., betweenness or degree centrality, to select the most useful seeds for offloading the cellular network. A similar social-aware approach is also used in~\cite{2014-infocom-toss,2013-mass-wu-offloading}. It is assumed in~\cite{2012-pmc-whitbeck-PT} that a centralised entity keeps tracks of the speed of the content dissemination process to decide when the cellular network should directly transmit the content to the interested users to guarantee that the delivery deadline is met with high probability. An actor-critic learning framework is designed in~\cite{2015-comcom-valerio} to understand when and to how many users the cellular network should directly transmit the content. In this context, an important area of research is the design of scalable and efficient network-aided offloading schemes, where the cellular network guides the mobile users in the connectivity management and dissemination phase. Finally, offloading data traffic when the requests for popular content are not synchronised is still an open issue~\cite{2014-cartoon-bruno}.

%
%
\section{Summary}\label{Sec:Summary}
\noindent
In this paper we have described the main scenarios, novel functionalities and mechanisms that will characterise the use, operation and performance of next-generation WLANs and provided an extensive and thorough review of the IEEE 802.11ac, IEEE 802.11aa, IEEE 802.11ah, IEEE 802.11p and IEEE 802.11af amendments. The paper also provides an up-to-date survey of the most representative work in this research area, summarising the key contributions to the current status and future evolution of WLANs. Differently from other 802.11-related surveys, this overview is structured with regards to the emerging WLAN application scenarios, such as M2M, cognitive radios and high-definition multimedia delivery. We also describe some open challenges that require further research in coming years~\cite{2014-comcom-future}, with special focus on software-defined MACs and the internet-working with cellular systems.
%
%
%
\section*{Acknowledgements}
\noindent
We would like to thank the anonymous reviewers for their insightful comments on the paper, as these comments led us to an improvement of the work. A special thank also to Toke H{\o}iland-J{\o}rgensen and Ognjen Dobrijevi{\'{c}} for the time they spent reviewing the manuscript. 
%
%
%
%
\section*{References}

\bibliographystyle{unsrt}
\bibliography{Bib80211aa,Bib80211ac,Bib80211p,Bib80211ah,Bib80211af,Bib80211ax,Bib80211program,Bib80211_offloading,Bib82011_hwexp,WLANSurveyBib}


\end{document}